\documentclass[11pt]{article}
\pdfoutput=1

\usepackage[]{Packages}
\bibliography{T-duality}

\usepackage{Definitions}

\usepackage{textcomp}

\preprint{\texttt{CERN-PH-TH-2015-007}\\
  \texttt{IPhT-t15/007}}

\newcommand{\OfficialTitle}{The braneology of 3D dualities}

\title{\vspace{2cm}
  {\Huge\textbf{\dosserif\OfficialTitle}}
}

\hypersetup{pdfauthor={Antonio Amariti and Davide Forcella and Claudius Klare and Domenico Orlando and Susanne Reffert},pdftitle={\OfficialTitle}}

\author{%
  \begin{minipage}{.8\linewidth}
    \vspace{1cm}
    \begin{center} \dosserif
      {\small 
        \textbf{Antonio~Amariti}\textsuperscript{$\spadesuit$},
        \textbf{Davide~Forcella}\textsuperscript{$\spadesuit\spadesuit$}, 
        \textbf{Claudius~Klare}\textsuperscript{$\heartsuit$}, 
        \textbf{Domenico~Orlando}\textsuperscript{$\spadesuit,\clubsuit$} and 
        \textbf{Susanne~Reffert}\textsuperscript{$\diamondsuit$}}
    \end{center}
    \vspace{1cm}
       \authorBlock{$\spadesuit$}{\textsc{lptens} -- \textsc{umr cnrs} 8549, 24, rue Lhomond, F-75231 Paris, France} 
    \authorBlock{$\spadesuit\spadesuit$}{Physique Th\'eorique et Math\'ematique and International Solvay Institutes\\ \textsc{ulb}, C.P. 231, 1050 Bruxelles, Belgium}
    \authorBlock{$\heartsuit$}{\textsc{ipht}, \textsc{cea}/Saclay, F-91191 Gif-sur-Yvette, France}
    \authorBlock{$\heartsuit$}{\textsc{ihes}, 35, Route de Chartres, F-91440 Bures-sur-Yvette, France}
    \authorBlock{$\clubsuit$}{\textsc{ipt} Ph. Meyer, 24, rue Lhomond, 75231 Paris, France}
    \authorBlock{$\diamondsuit$}{\textsc{ph -- th}, \textsc{cern}, CH-1211 Geneva 23, Switzerland}
    \authorBlock{$\diamondsuit$}{\textsc{itp -- aec}, University of Bern, Sidlerstrasse 5, CH-3012 Bern, Switzerland}
  \end{minipage}
}

\date{}

\begin{document}

\setstretch{1.15}

\numberwithin{equation}{section}

\begin{titlepage}

  \newgeometry{top=23.1mm,bottom=46.1mm,left=34.6mm,right=34.6mm}

  \maketitle

  \thispagestyle{empty}

  \vfill\dosserif
  \vspace{1cm}
  \abstract{\normalfont \noindent
      In this paper we study the reduction of four-dimensional Seiberg
      duality to three dimensions from a brane perspective.\\
      We reproduce the non-perturbative dynamics of the three-dimensional  field theory
      via a T--duality at finite radius and the action of Euclidean D--strings.
      In this way we also overcome certain  issues regarding the 
      brane description of Aharony duality.
      Moreover we apply our strategy to more general dualities, such as toric duality for 
      \M2--branes and dualities with adjoint matter fields.
    }

  \vfill

\end{titlepage}

\restoregeometry


\section{Introduction}

Gaining a better understanding of the \textsc{ir} properties of strongly coupled field theories is an ambitious, yet non-trivial goal. One approach which has been successfully applied many times over the last decades is duality. On the two sides of a duality, different degrees of freedom are used to describe the same physical phenomenon. One of the main strengths of this idea is that strongly coupled theories can have a weakly coupled counterpart which is accessible via perturbative analysis. It has however been difficult to find examples of such dualities. An exception are supersymmetric theories, where a plethora of dualities has been formulated. 

In the minimal case of four supercharges in 3+1 space-time dimensions, Seiberg duality~\cite{Seiberg:1994pq} has played a major role in the investigation of field theory dynamics. Many checks and applications have since followed. 
This duality, originally formulated for $SU(N_c)$  \textsc{sqcd} with $N_f$ fundamentals and anti-fundamentals, is a quite general property of 3+1-dimensional supersymmetric gauge theories. It has been extended to cases with real gauge groups and more complicated matter content (see for example~\cite{Intriligator:1995id,Intriligator:1995ne,Kutasov:1995ss}).
Also cases with product groups have been studied, and it was realized that this duality plays an important role in the study of \textsc{rg} flows in the AdS/CFT correspondence~\cite{Klebanov:2000hb}.
Another extension of this duality is to 2+1 dimensions. Some early attempts were made in the 90s~\cite{Intriligator:1996ex,deBoer:1996mp,
deBoer:1996ck,deBoer:1997ka,Aharony:1997bx,Aharony:1997gp,Karch:1997ux}, but the 
2+1--dimensional version of Seiberg duality has become more popular only in recent years.

There are two main reasons for this renewed interest.
On the one hand, localization has allowed the exact computation of the partition function of theories that preserve some supercharges in curved spaces. 
One the other hand,
the discovery of the \ac{abjm} model~\cite{Aharony:2008ug} has extended the AdS/CFT duality to three dimensions. The simplest case of 
AdS$_4 \times S^7/Z_k$ has been found to be dual to a \ac{cs} quiver gauge theory with $\mathcal{N}=6$ supersymmetry.
After the \ac{abjm} model, the correspondence has been studied also for cases with less supersymmetry.
Many theories have been conjectured to describe 
the \textsc{ir} dynamics of \M2--branes probing a Calabi--Yau
cone $X = C(Y )$ over the seven-dimensional Sasaki--Einstein manifold $Y$~\cite{Hanany:2008cd,Martelli:2008si,Hanany:2008fj}.
They preserve generally $\mathcal{N}=2$ supersymmetry, the same number of supercharges as in $\mathcal{N}=1$ in 3+1 dimensions.

Motivated by these observations, many extensions of the duality in the 2+1--dimensional $\mathcal{N}=2$ case
have been proposed and analyzed~\cite{Aharony:2008gk,Giveon:2008zn,Niarchos:2008jb,Amariti:2009rb,Benini:2011mf,Kapustin:2011vz,Kim:2013cma,Aharony:2014uya}.
Examples have been found and it was observed that instead of a single duality like in four dimensions, there are many possibilities. 
These are often close to the four-dimensional parent duality but differ from each other in some aspects.
There are essentially two reasons for this proliferation of dualities in 2+1~dimensions:
\begin{itemize}
\item the first is that in three dimensions it is possible to write a
  topological \ac{cs} action for the gauge field, which affects the \textsc{ir} dynamics quite strongly;
\item secondly, the moduli space has a complicated structure due to the presence of a real scalar field in the vector multiplet, coming from the reduction  of the last component of the 3+1-dimensional gauge field.
\end{itemize}
This scalar moreover allows the existence of real (non-holomorphic)  mass terms in the action.
The absence of anomalies  in 2+1 dimensions allows also the existence of chiral\footnote{Note that even if the notion of chirality is
absent in three dimensions, by ``chiral'' we refer to the chirality the four-dimensional parent theory.}
real masses and \ac{cs} terms which are very closely connected. By integrating out a certain amount of matter with non-vanishing real mass, one can generate a \ac{cs} action. This mechanism is crucial in the study of three-dimensional dualities.

\bigskip
It is natural to wonder whether the various three-dimensional dualities are somehow related and whether they can be classified in a unified way.
At first sight there are analogies between these dualities and the four-dimensional case. Many of these relations are more evident from a stringy perspective.
An interesting starting point to classify and possibly find new dualities in three-dimensional $\mathcal{N}=2$ gauge theories consists in compactifying the four-dimensional dualities on a circle 
at finite radius~\cite{Seiberg:1996nz}. By shrinking this circle one expects to obtain a duality in three dimensions.
This idea has been deeply investigated in~\cite{Aharony:2013dha}, where a non-canonical dimensional reduction was 
performed.
By naive dimensional reduction of two dual phases one obtains a new duality that is valid 
only below a too small energy scale. This is because the holomorphic scale associated to the gauge coupling decreases with  the radius of the circle. There is a different way to consider this limit, by interpreting the 
theory on the finite circle as an effective three-dimensional theory. When the low-energy spectrum of the duality on the circle is considered, the two phases remain dual thanks to the non-perturbative effects 
generated by the circle itself.
This idea as been applied in~\cite{Aharony:2013dha} to \textsc{sqcd} with flavor in the fundamental representation 
of unitary symplectic groups
and in~\cite{Aharony:2013kma} for the orthogonal case\footnote{See also~\cite{Niarchos:2012ah} for an earlier attempt.}.
New dualities have been found in this way, and it has been shown that from these one can 
also flow to previously known examples (see \emph{e.g.}~\cite{Aharony:1997gp} 
and~\cite{Giveon:2008zn}).
The procedure is more general and can be applied to other configurations. The problem 
is that the non-perturbative structure  of the theories at finite radius strongly depends  on the details of the gauge and matter content. The counting of the zero modes and the presence of a compact moduli space in the theory on the
circle can strongly affect the non-perturbative dynamics, so one has to study the different cases separately. For example, the case with adjoint matter and unitary groups has been discussed 
in~\cite{Nii:2014jsa},
while the analysis of the s--confining case was started in~\cite{Csaki:2014cwa}.

\bigskip

In this paper we discuss the mechanism to generate three-dimensional dualities from four dimensions based on the brane description of the associated field theories. We also
make progress towards the geometric description of the duality of Aharony~\cite{Aharony:1997gp}. 
The string realization is a \textsc{uv} completion of the gauge theories. One might wonder if the resulting extra massive modes spoil a duality proof based on the brane construction. This is not the case: the matching of points in moduli space (\emph{i.e.} the equivalence of gauge theories) is protected by supersymmetry;  only the form of the Kähler potential depends on the details of the string \textsc{uv} completion.

We start by considering the four-dimensional representation of Seiberg duality as seen from a brane
setup. This can be represented as a \tIIA system, with $N_c$ \D4--branes stretched between two non-parallel \NS5--branes and $N_f$ \D6--branes. This system reproduces the low-energy spectrum of $\mathcal{N}=1$ $SU(N_c)$ \textsc{sqcd} 
with $N_f$ fundamentals and anti-fundamentals in four dimensions. Seiberg duality is obtained by exchanging the two \NS{}--branes. The reduction of this duality in terms of branes can be sketched as follows. 
First one compactifies one space-like direction of the spacetime, for example $x_3$. 
Then a \emph{T--duality} is performed along this compact direction. This duality 
transforms the \D{p}--branes considered above into \D{(p-1)}--branes, leaving the \NS{}--branes unchanged.
If the T--dual radius is large enough we can consider the theory as effectively three-dimensional.
The non-perturbative dynamics, coming from the presence of the circle,
is captured at the brane level by \D1--strings stretched between the \D3 and the \NS{}--branes.
This  analysis reproduces the field theory results found in~\cite{Aharony:2013dha}
in terms of brane dynamics.
In field theory, by adding some mass deformation to this duality, one obtains Aharony duality.
Here we reproduce this \textsc{rg} flow from the brane perspective, where many aspects of this flow 
have a simple physical interpretation.

The brane analysis allows us to study, in a general framework,
theories with a more complicated matter content (\emph{e.g.} tensor matter) and gauge groups (orthogonal and symplectic). 
Also in these cases Seiberg duality corresponds to the exchange of the \NS{}--branes. The reduction 
of these theories to three dimensions follows from the steps discussed above.
Indeed we (re)-obtain the reduction discussed in~\cite{Nii:2014jsa} for the duality with adjoint matter by using the brane representation of this theory.
We also propose the reduction of the duality for quiver gauge 
theories and discuss the relation of the three-dimensional dualities obtained by dimensional
reduction and the dualities obtained in~\cite{Amariti:2009rb} for \M2--branes on CY fourfolds.
We conclude by proposing the extension of our procedure to the case with tensor matter and real gauge groups, higher supersymmetry and lower dimensionality.

\section{Mini review of known results}
\subsection{Field theory reduction of 4D dualities to 3D}

In this section we review the relevant aspects of three-dimensional field theories and of the dimensional reduction
of the four-dimensional Seiberg duality for $U(N_c)$ \textsc{sqcd} with $N_f$ flavors to three dimensions.

\paragraph{Some aspects of $\mathcal{N}=2$ three-dimensional field theories.}
Three-dimensional theories with four supercharges have an additional (real) scalar $\sigma$ in the vector multiplet with respect to the four-dimensional case.
Classically, this scalar implies the existence of a Coulomb branch parametrized by the \ac{vev}~$\langle \sigma \rangle$, which generically breaks the rank $r$ gauge group $G$ to $U(1)^r$.
Another feature of three-dimensional gauge theory is that a $U(1)$ gauge field $A_\mu$ can be dualized to a scalar $\phi = \di \st F$, where $F$ is the $U(1)$ field strength.
On the Coulomb branch we have $r$ such dual photons, one from each $U(1)$ factor.
We can dualize the $i$th vector multiplet to a chiral one with lowest component $Y_i \equiv e^{i \phi_i + \sigma_i/e_3^2}$, where \(e_3\) is the gauge coupling.

Quantum corrections can lift some of the directions on the Coulomb branch.
For $U(N_c)$ \textsc{sqcd} with $N_f > N_c $ only two directions $Y \equiv Y_1$ and $\widetilde Y \equiv Y_{N_c}$ remain unlifted~\cite{Aharony:1997bx}.
The Coulomb branch coordinates $Y, \widetilde Y$  have a UV interpretation as monopole operators in the field theory~\cite{Aharony:1997bx},
which are excitations of magnetic flux $(\pm 1, 0,\dots,0)$ in the Cartan subgroup $U(1)^r$.
Note that $Y, \widetilde Y$ are charged under the topological symmetry $U(1)_J$, that shifts the dual photon.

\bigskip

\paragraph{Duality at finite radius.}

Let us consider a $U(N_c)$ gauge theory with $N_f$
fundamentals and anti-fundamentals $Q$ and $\widetilde Q$ without superpotential as the four-dimensional electric theory. For $N_f > N_c + 1$ the theory admits a dual description\footnote{The
confining case \(N_f = N_c+1\) was discussed recently in~\cite{Csaki:2014cwa}.} (the so-called Seiberg-dual)
in terms of a 
magnetic theory with gauge group $U(N_f-N_c)$, $N_f$ fundamentals and anti-fundamentals
$q$ and $\tilde q$, a meson $M=Q \widetilde Q$ and a superpotential $W=M\, q \tilde q$.

If we put both the electric and the magnetic theory on $\mathbb{R}^3 \times S^1$ with finite circle radius $R_3$, a three-dimensional description is obtained by keeping the scales $\Lambda$, $\widetilde \Lambda$ and the radius $R_3$ fixed and by looking at  energies $E \ll \Lambda, \widetilde \Lambda, 1/R_3$.
In this limit the dynamics is effectively three-dimensional and, as discussed in~\cite{Aharony:2013dha},
the $4$D duality reduces to a new duality in $3$D. 
Crucially, we can still see the effect of the finite circle radius through the non-perturbative superpotentials
\begin{align}
  \label{eq:etasup}
  W_{\eta} &= \eta\, Y \widetilde Y, &  W_{\eta'} &= \eta' y \tilde y,
\end{align}
where $y, \tilde y$ are the monopoles of the magnetic theory
and $\eta=e^{-8 \pi/ (R_3 e_3^2)} = \Lambda^{2b}$ (recall that $2 \pi R_3 e_3^2=  e_4^2$).
The power $b$ is the coefficient of the one-loop \ac{nsvz} $\beta$-function; analogous definitions hold for the magnetic $\eta'$.

The superpotentials~$\eqref{eq:etasup}$ completely lift the Coulomb branch. 
Note also that~\eqref{eq:etasup} break the axial symmetry $U(1)_A$ which in four dimensions is broken by anomalies.
The global symmetries coincide with the ones of the four-dimensional parent theory. 
The charges of the $3$D fields are given in
Table~\ref{tab:global-symmetries}.
\begin{table}
  \centering
  \begin{tabular}{cccccc}
    \toprule
                     & \(SU(N_f)_L\)      & \(SU(N_f)_R\)      & \(U(1)_R\)                 & \(U(1)_J\) \\
    \midrule
    \(Q\)            & \(N_f\)            & \(1\)              & \(\Delta\)               & \(0\)      \\
    \(\widetilde Q\) & \(1\)              & \(\overline{N_f}\) & \(\Delta\)               & \(0\)      \\ 
    \midrule
    \(q\)            & \(\overline{N_f}\) & \(1\)              & \(1-\Delta\)             & \(0\)      \\
    \(\widetilde q\) & \(1\)              & \(N_f\)            & \(1-\Delta\)             & \(0\)      \\
    \(M\)            & \(N_f\)            & \(\overline{N_f}\) & \(2 \Delta\)             & \(0\)      \\
    \(Y\)            & \(1\)              & \(1\)              & \(N_f(1-\Delta)-N_c+1\)  & \(1\)      \\
    \(\widetilde Y\) & \(1\)              & \(1\)              & \(N_f(1-\Delta)-N_c+1\)  & \(-1\)     \\
    \(y\)            & \(1\)              & \(1\)              & \(-N_f(1-\Delta)+N_c+1\) & \(1\)      \\
    \(\tilde y\) & \(1\)              & \(1\)              & \(-N_f(1-\Delta)+N_c+1\) & \(-1\)     \\ \bottomrule
  \end{tabular}
  \caption{Global symmetries of the three-dimensional fields in the reduction of Seiberg duality. Note that the monopoles are charged under $U(1)_R$ because of quantum corrections~\cite{Benna:2009xd,Jafferis:2009th,Benini:2011cma}.}
  \label{tab:global-symmetries}
\end{table}

\paragraph{Aharony duality.}

If we want to recover a more \emph{classic} three-dimensional duality, we have to integrate out some matter fields.
 If we consider for example $N_f+2$ flavors and integrate out two pairs of fundamentals and anti-fundamentals, one with positive large mass and one with opposite large mass, we obtain a $U(N_c)$ gauge theory with $N_f$ flavors.
In the magnetic theory the dual quarks and the mesons acquire large masses fixed by their charges with respect to the global symmetries.

To preserve the duality one has to higgs the gauge symmetry by
assigning a large \ac{vev} to some components of the scalar 
in the vector multiplet~\cite{Aharony:2013dha}. The \ac{vev} 
breaks the gauge symmetry as  $U(N_f+2-N_c)\rightarrow U(N_f-N_c) \times U(1)^2$.
Each $U(1)$ sector has a fundamental and an anti-fundamental field and a singlet.
The three gauge sectors are coupled via \ac{ahw} superpotentials.
The superpotential of this dual theory is given by
\begin{equation}
  \label{eq:W-preAharony}
  W_{\text{m}} = M\, q \tilde q + M_1 q_1 \tilde q_1+ M_2 q_2 \tilde q_2 + y \tilde y_1
+ \tilde y y_2 + \eta' y_1 \tilde y_2.
\end{equation} 
 The $U(1)$ sectors can be dualized to sectors containing only 
 singlets~\cite{Intriligator:1996ex,Aharony:1997bx,deBoer:1997ka}. 
Observe that in absence of  the singlets $M_i$ each mirror sector coincides with an XYZ
model~\cite{Aharony:1997bx}, 
which is a Wess--Zumino model with three chiral fields $X$, $Y$ and $Z$ and superpotential $W= XYZ$. Here the dual mesons $q_i \tilde q_i$ become massive because of the superpotential. 
Moreover, the term $y_1 \tilde y_2$
is a mass term in the mirror theory, because $y_{i}, \tilde y_{i}$ are singlets.
In the limit $E\ll \Lambda,\widetilde \Lambda,1/R_3$ we can integrate out this mass term.
 Finally the superpotential in Eq.~\eqref{eq:W-preAharony} becomes
 \begin{equation} \label{eq:Aharony}
   W_{\text{m}} = M q \tilde q + y Y +\tilde y \widetilde Y,
\end{equation} 
where the identifications  $Y = \tilde y_1$ and $\widetilde Y = y_2$ follow from the quantum charges of the singlets.
The superpotential (\ref{eq:Aharony}) reproduces the one expected from the duality of Aharony.

\subsection{Non-perturbative superpotentials from the brane picture}
\label{sec:superpotential-from-branes}

In this section we review some aspects of the generation of the 
non-perturbative superpotential for \textsc{sqcd} at finite radius from the brane perspective.
The discussion will be relevant in the next section when studying the reduction of
four-dimensional Seiberg duality to three dimensions.

\paragraph{Super Yang--Mills.}
Consider $\mathcal{N}=1$ \ac{sym}. The theory  is described by an \NS5--brane, an \NS5'--brane and $N_c$ \D4--branes extended as shown in Table~\ref{tab:IIA-sqcd-no-flavors}. The \D4s are suspended between the \NS5 and the \NS5'. The four-dimensional gauge coupling is \(e_4^2 = g_4^2/\ell_6 =  (2\pi)^2 \sqrt{\alpha'} / \ell_6 \) where \(\ell_6 \) is the distance between the \NS5--branes and \(g_4^2 = (2\pi)^2 \sqrt{\alpha'}\) is the \D{}--brane coupling constant.
\begin{table}
  \centering
  \begin{tabular}{lcccccccccc}
    \toprule
     & 0 & 1 & 2 & 3 & 4 & 5 & 6 & 7 & 8 & 9\\
    \midrule
    NS & $\times$ & $\times$ & $\times$ & $\times$ & $\times$ & $\times$ &  &  & \\
    NS' & $\times$ & $\times$ & $\times$ & $\times$ &  &  &  &  & $\times$ & $\times$\\
    \D4 & $\times$ & $\times$ & $\times$ & $\times$ &  &  & $\times$ &  & \\ \bottomrule
  \end{tabular}
  \caption{Brane configuration for \(\mathcal{N}=1, d =4 \) \ac{sym}. The \D4--branes are suspended between the two \NS5s.}
  \label{tab:IIA-sqcd-no-flavors}
\end{table}

We consider compact \(x_3 \sim x_3 + 2 \pi R_3\) and perform a T--duality along that direction.
In the resulting \tIIB frame the \D4s have turned into \D3--branes while the \NS{} and \NS'--branes remain unchanged.
This setup describes $U(N_c)$ \ac{sym} on $\setR^3 \times S^1$.
The theory has the entire Coulomb branch lifted and  $N_c$ isolated vacua~\cite{Witten:1982df}.
The isolated vacua correspond to stable supersymmetric configurations of the brane system.
As we will review below, in absence of \D5--branes there is a repulsive force between the \D3s.
A stable brane configuration corresponds to distributing the \D3--branes along $x_3$ at equal distances. 
Here all moduli are lifted as the \D3--branes cannot move freely due to the repulsive force.

\begin{figure}
  \begin{center}
    \begin{tikzpicture}
    \begin{scope}[shift={(-3,-2)}]
      \begin{footnotesize}
      \draw[->] (-0.2,0) arc (-140:135:.2 and .4);
      \draw[->] (-0.4,.25) -- (.6,.25);
      \draw (0.6,.45) node[]{\(x_6\)};
      \draw (0.1,.9) node[]{\(x_3\)};
      \end{footnotesize}
    \end{scope}

      \begin{scope}[thick]
        \begin{footnotesize}
          \draw[fill=black!10,black!10] (0,-.5) rectangle (7,.5); 
          \draw (0,-.5) -- (7,-.5);
          \draw (0,.5) -- (7,.5);
          \draw (.5,1.5) -- (6.7, 1.5);
          \draw (.5,-1.5) -- (6.7, -1.5);

          \draw[fill=white] (0,0) ellipse (1 and 2);
          \draw[dashed] (6,0) ellipse (1 and 2);

          \draw (0,2) -- (6,2);
          \draw (0,-2) -- (6,-2);

          \draw[fill=white] (6.9,.5) circle (5pt);
          \draw (6.9,.5) node[cross=4pt]{};
          \draw[fill=white] (6.9,-.5) circle (5pt);
          \draw (6.9,-.5) node[cross=4pt]{};

          \draw[fill=white] (6,2) circle (5pt);
          \draw (6,2) node[cross=4pt]{};
          \draw[fill=white] (6,-2) circle (5pt);
          \draw (6,-2) node[cross=4pt]{};

          \draw[fill=white] (6.7,1.5) circle (5pt);
          \draw (6.7,1.5) node[cross=4pt]{};
          \draw[fill=white] (6.7,-1.5) circle (5pt);
          \draw (6.7,-1.5) node[cross=4pt]{};

          \draw[fill=white] (5.1,.8) circle (5pt);
          \draw (5.1,.8) node[cross=4pt]{};
          \draw[fill=white] (5.1,-.8) circle (5pt);
          \draw (5.1,-.8) node[cross=4pt]{};

          \draw (-.8,-2) node[]{\NS5};
          \draw (6.7,-2) node[]{\NS5'};
          \draw (3,-1) node[]{\D3};         
          \draw (4,0) node[]{\D1};
          \draw (7.4,.5) node[]{\D5};

        \end{footnotesize}
      \end{scope}
    \end{tikzpicture}
  \end{center}
  \caption{\D1--branes (in grey) stretched between \D3--branes with
    compact \(x_3\). The \D5--branes are represented by \(\otimes\) symbols, the \NS5 by a continuous line and the \NS5' by a dashed line.}
  \label{fig:D1-between-D3}
\end{figure}
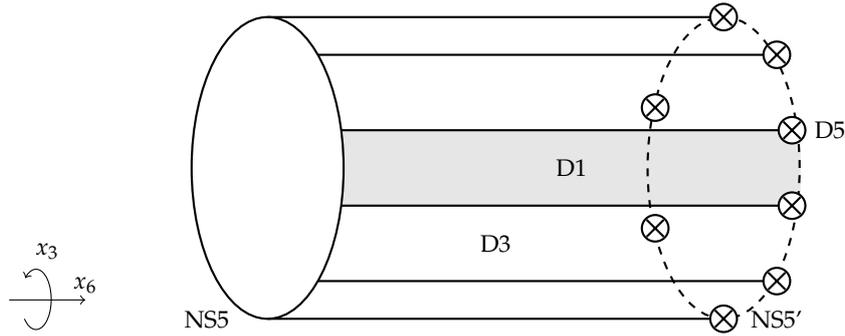

The repulsive force is a non-perturbative quantum effect. 
From the $3$D field theory point of view, a non-perturbative superpotential induced by three-dimensional instantons is generated.
From the $4$D perspective these instantons are monopole configurations.
In the brane picture these monopoles are represented by Euclidean \D1--strings stretched between 
each pair of \D3--branes and the \NS{} and \NS'--branes, as depicted in figure~\ref{fig:D1-between-D3} 
as shaded area along $x_6$ and $x_3$ respectively. 
The contribution of the monopoles can be computed following~\cite{deBoer:1997ka,Hanany:1996ie} as $e^{-S}$, where $S$ is the \D1 world-sheet action.
This action has two pieces, the Nambu--Goto action and a contribution from the boundary of the \D1s.
The Nambu--Goto term is proportional to the area of the \D1 and involves the scalar $\sigma$ that parametrizes the position of the \D3.
The boundary term is proportional to the dual photon $\phi$.
By combining everything together one obtains the monopole contribution as
\begin{equation}
\label{WD1}
W = \sum_{i=1}^{N_c-1} e^{\Sigma_i-\Sigma_{i+1}},
\end{equation}
where $\Sigma_i = \sigma_i/e_3^2+i \phi_i $ and $e_3^2 = (\sqrt{\alpha'}/R_3)
(g_3^2 /\ell_6) = 2 \pi \sqrt{\alpha'}/ (R_3 \ell_6)$ is the
three-dimensional gauge coupling\footnote{Here \(g_3^2 = 2 \pi \), and \( \sqrt{\alpha'}/R_3\) is the contribution of the \tIIB dilaton so that \(e_4^2/ e_3^2 = 2 \pi R_3\).}.       
Observe that the result is expressed in terms of the Coulomb branch coordinates, 
in terms of operators one can associate $e^{\Sigma_i}$ to the monopole operators.

As is easily seen from the brane picture, for $x^3$ being compact and finite there is another contribution from \D1--branes stretching from the $N_c$th to the $1$st \D3--branes,
which sit at positions $\sigma_{N_c}$ and $\sigma_1 + 2 \pi R_3$ respectively. 
By following the calculation of~\cite{Davies:1999uw,Katz:1996th} one finds
\begin{equation}
  \label{etaWW}
  W_\eta = \eta \, e^{\Sigma_{N_c}-\Sigma_1}  ,
\end{equation}
where $\eta$ incorporates the radius dependence from the position of the $1$st
\D3--brane.
This is the $\eta$--superpotential $\eqref{eq:etasup}$ which proves important in the reduction of $4$D to $3$D dualities.

\paragraph{Fundamental matter.}
The picture becomes more interesting when we introduce matter fields.
Here we consider only the case $N_f>N_c$ as we are interested in the dimensional reduction of Seiberg duality.
As mentioned above two directions of the moduli space remain unlifted, which can be seen from the brane picture as follows.
In the \tIIA frame fundamental matter is associated to $N_f$ \D6--branes extended along $0123789$ and sitting on the \NS'--brane. In the T--dual
frame they become \D5--branes. Strings between the stack of \D3--branes and the \D5s correspond to 
$N_f$ massless fundamentals $Q$ and anti-fundamentals $\widetilde Q$.

When \D5--branes sitting at $x_3 =0$ intersect the worldsheet of the \D1--strings,
they contribute two additional zero modes to the \D1--instanton and the superpotential in Eq.~(\ref{WD1}) is not generated.
In this sense the \D5--branes screen the repulsive force between the \D3--branes~\cite{deBoer:1997kr,Elitzur:1997hc}.
One finds that the screening happens for the $1$st and the $N_c$th \D3--brane, which are still free to move without being subjected to any force.
There is however the interaction given by superpotential Eq.~(\ref{etaWW}), which lifts one modulus. 

For $x_3$ being non-compact, there is no superpotential $\eqref{etaWW}$ and we remain with a two-dimensional moduli space 
corresponding to the forceless motion of the $1$st and the $N_c$th \D3--brane.

\section{The braneology of the reduction}
\label{sec:Braneology}
In this section we study the reduction of four-dimensional Seiberg duality for \textsc{sqcd} 
to three dimensions from the perspective of brane dynamics.
Note that we consider the four-dimensional gauge symmetry to be $U(N_c)$ rather than $SU(N_c)$. 
In field theory this enhancement is obtained by gauging the baryonic symmetry. 

We start by considering a stack of $N_c$ \D4--branes, one \NS5--brane, one \NS5'--brane and $N_f$ \D6--branes. In this \tIIA description the branes are extended as shown in Table~\ref{tab:IIA-sqcd}.
\begin{table}
  \centering
  \begin{tabular}{lcccccccccc}
    \toprule  & 0 & 1 & 2 & 3 & 4 & 5 & 6 & 7 & 8 & 9\\
    \midrule
    NS & $\times$ & $\times$ & $\times$ & $\times$ & $\times$ & $\times$ &  &  & \\
    NS' & $\times$ & $\times$ & $\times$ & $\times$ &  &  &  &  & $\times$ & $\times$\\
    \D4 & $\times$ & $\times$ & $\times$ & $\times$ &  &  & $\times$ &  & \\
    \D6 & $\times$ & $\times$ & $\times$ & $\times$ &  &  &  & $\times$ & $\times$ & $\times$\\ \bottomrule
  \end{tabular}
  \caption{Brane configuration for \(\mathcal{N}=1, d =4 \) \textsc{sqcd} with \(N_f \) flavors. The \(N_c\) \D4--branes are extended between the \NS5 and the \(N_f\) \D6--branes sit on the \NS5'.}
  \label{tab:IIA-sqcd}
\end{table}
The \D4--branes are suspended between the \NS5--branes.
The distance between the \NS5--branes is proportional to the inverse gauge coupling of the four-dimensional theory.
There are two possible configurations: in the first (electric) the \NS5 is on the left and the \NS5' on the right with \(N_c\) \D4--brane in between.
Moving the \NS5 to the right we obtain the second configuration (magnetic) with \((N_f - N_c)\) suspended \D4--branes.
The next step consists of compactifying both dual phases along $x_3$ and studying the 
two dual theories at finite radius.

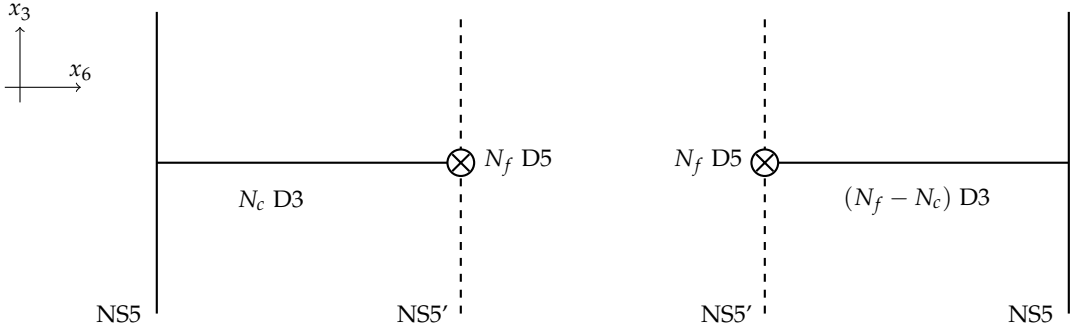
\begin{figure}
  \begin{center}
    \begin{tikzpicture}
      \begin{scope}[shift={(-2,1)}]
        \begin{footnotesize}
          \draw[->] (0,3) -- (1,3);
          \draw[->] (.2,2.8) -- (.2,3.8);
          \draw (1,3.2) node[]{\(x_6\)};
          \draw (.2,4) node[]{\(x_3\)};
        \end{footnotesize}
      \end{scope}

      \begin{scope}[thick]
        \begin{footnotesize}
          \draw (0,1) -- (0,5);
          \draw (-0.5,1) node[]{\NS5};
          \draw (0,3) -- (4,3);
          \draw (1.5,2.5) node[]{\(N_c\) \D3};
          \draw[dashed] (4,1) -- (4,5);
          \draw (3.5,1) node[]{\NS5'};
          \draw[fill=white] (4,3) circle (5pt);
          \draw (4,3) node[cross=4pt]{};
          \draw (4.75,3) node[]{\(N_f\) \D5};
        \end{footnotesize}
      \end{scope}
      \begin{scope}[shift={(8,0)},thick]
        \begin{footnotesize}
          \draw (4,1) -- (4,5);
          \draw (3.5, 1) node[]{\NS5};
          \draw (0,3) -- (4,3);
          \draw (2,2.5) node[]{\((N_f - N_c)\) \D3};
          \draw[dashed] (0,1) -- (0,5);
          \draw (-.5,1 ) node[]{\NS5'};
          \draw[fill=white] (0,3) circle (5pt);
          \draw (0,3) node[cross=4pt]{};
          \draw (-.75,3) node[]{\(N_f\) \D5};
        \end{footnotesize}
      \end{scope}
    \end{tikzpicture}
  \end{center}
  \caption{Brane description of three-dimensional Seiberg duality. From the brane description one can 
    see that the electric superpotential is $W_e=0$ and the magnetic one is $W_{\text{m}}=M q \tilde q$.}
  \label{fig:Seiberg-duality-D4D6}
\end{figure}

\subsection{Duality at finite radius}

If we consider $x_3$ to be compact we can perform a T--duality and obtain an 
effective three-dimensional $\mathcal{N}=2$
theory. In this case the \D4 and the \D6--branes become \D3s and \D5s respectively, while the
\NS{} and \NS'--branes are left unchanged. In Figure~\ref{fig:Seiberg-duality-D4D6}
the electric and magnetic Seiberg--dual theories are shown, where the horizontal direction is $x_6$ and the vertical one is $x_3$.
In the \tIIB description the branes are extended as shown in Table~\ref{tab:IIB-sqcd}.
This is a $U(N_c)$ gauge theory with $N_f$ fundamentals and anti-fundamentals as discussed in Section~\ref{sec:superpotential-from-branes}.
The global symmetry group is $SU(N_f)_L \times SU(N_f)_R \times U(1)_R \times U(1)_J$.
The axial symmetry $U(1)_A$ under which $Q$ and $\widetilde Q$ have the same charge is broken by the superpotential in Eq.~(\ref{etaWW}).
\begin{table}
  \centering 
  \begin{tabular}{lcccccccccc}
    \toprule  & 0 & 1 & 2 & 3 & 4 & 5 & 6 & 7 & 8 & 9\\
    \midrule
    NS & $\times$ & $\times$ & $\times$ & $\times$ & $\times$ & $\times$ &  &  & \\
    NS' & $\times$ & $\times$ & $\times$ & $\times$ &  &  &  &  & $\times$ & $\times$\\
    \D3 & $\times$ & $\times$ & $\times$ &  &  &  & $\times$ &  & \\
    \D5 & $\times$ & $\times$ & $\times$ &  &  &  &  & $\times$ & $\times$ & $\times$\\ \bottomrule
  \end{tabular}
  \caption{Brane configuration for \(\mathcal{N}=2, d =3 \) \textsc{sqcd}. The \(N_c\) \D3--branes are extended between the \NS5 and the \(N_f\) \D5--branes sit on the \NS5'.}
  \label{tab:IIB-sqcd}
\end{table}

Let us see how the global symmetries are realized in the brane picture.
\begin{itemize}
\item The brane system is invariant under rotations in the $(4, 5)$ and in the $(8, 9)$--plane.
The corresponding symmetry $U(1)_{45} \times U(1)_{89}$ is part of the Lorentz group in 9+1 dimensions and rotates the supercharges.
It is hence an \(R\)--symmetry\footnote{Recall
that in four dimensions, where we have a \emph{quantum} description of the field theory through M--theory, one can see that only one particular combination of $U(1)_{45}$ and $U(1)_{89}$
is a symmetry in the lift to $11$d (due to bending of the branes). This singles out the anomaly-free $R$--symmetry. 
Here we find both classical symmetries potentially realized in the field theory.}.
However, the axial $U(1)$ subgroup leaves the supercharges invariant while it rotates $Q$ and $\widetilde Q$ in the same way,
appearing as axial symmetry $U(1)_A$ in the field theory. 
While admissible in $3$D, on $\mathbb{R}^3 \times S^1$ it is broken as will be discussed in the next paragraph.
\item The non-Abelian flavor symmetry comes from the stack of $N_f$ \D5--branes.
The chiral nature comes from the fact that the branes can be broken at the intersection with the \NS5', leading to two semi-infinite stacks, $\D5_L$ and $\D5_R$, one extended in $x_7>0$ and the other in $x_7<0$.
The freedom to move the branes in each stack independently along $x_3$ signals the presence of two independent $SU(N_f)$ rotations~\cite{Brodie:1997sz}.
The flavor group is $U(N_f) \times U(N_f)$ but it turns out that one combination of the two $U(1)$ factors appears as baryonic and one as axial symmetry. 
Let us make this more precise. The $U(1) \subset U(N_f)$ real mass, \emph{i.e.} the \ac{vev} of the scalar in the corresponding $U(1)$ vector multiplet, 
corresponds to collectively shifting the stack of semi-infinite \D5s. 
The \emph{axial} $U(1)$ subgroup in $U(N_f) \times U(N_f)$ corresponds to moving the two stacks in opposite directions; 
the \emph{diagonal} subgroup corresponds to shifting both stacks (hence the original \D5--brane) together.
The latter is equivalent to a shift of the stack of the \D3--branes, 
which is a $U(1) \subset U(N_c)$ gauge transformation corresponding to the gauged baryonic symmetry.
The former affects the fundamentals and the anti-fundamentals as does the axial subgroup of $U(1)_{45} \times U(1)_{89}$ and is identified with turning on a real mass for $U(1)_A$.\\
On a circle (when $x_3$ is compact) the $U(1)_A$ is broken. In the brane picture 
this breaking can be visualized as follows. 
When moving the stack of $\D5_L$s in $x_3 < 0 $ and the stack of $\D5_R$s in $x_3 > 0$,
charge conservation requires the generation of a $(1,N_f)$ fivebrane along the directions $x_3$ and $x_7$~\cite{Aharony:1997ju,Kitao:1998mf} (see Figure~\ref{U(1)A}).
The \NS{}--brane now cannot close anymore on the circle without breaking supersymmetry.
This obstruction precludes the realization of the $U(1)_A$ symmetry in the brane setup.
This explains how the compactness of $x_3$ geometrically
reproduces the role of the $\eta$-superpotential in breaking the $U(1)_A$ symmetry\footnote{This mechanism is in spirit similar to the breaking of the axial symmetry in four-dimensional gauge theories. In fact, if we T--dualize and lift our configuration to M--theory, the system of \D5s and \NS' is lifted to a single \M5--brane wrapped on a Riemann surface (\emph{brane bending}).}.
\begin{figure}
  \centering
  \begin{tikzpicture}
    \begin{scope}[shift={(-2,-2)}]
      \begin{footnotesize}
      \draw[->] (-0.2,0) arc (-140:135:.2 and .4);
      \draw[->] (-0.4,.25) -- (.6,.25);
      \draw (0.6,.45) node[]{\(x_7\)};
      \draw (0.1,.9) node[]{\(x_3\)};
      \end{footnotesize}
    \end{scope}
    \begin{scope}[thick]
      \begin{footnotesize}
        \draw[thin] (0,0) ellipse (1 and 2);
        \draw[thin] (4,2) arc (90:-90:1 and 2);
        \draw[thin] (0,2) -- (4,2);
        \draw[thin] (0,-2) -- (4,-2);

        \draw (2.5,2) arc (90:-90:1 and 2);
        \draw[dashed] (2.5,-2) arc (-90:-270:1 and 2);

        \draw (1,0) -- (5,0);

        \draw (2.4,.4) node[]{\(N_f\) \(\D5_L\)};
        \draw (4.2,-.4) node[]{\(N_f\) \(\D5_R\)};

        \draw (3.8,1.2) node[]{\NS5'};
      \end{footnotesize}
    \end{scope}

    \begin{scope}[shift={(6.5,0)},thick]
      \begin{footnotesize}
        \draw[thin] (0,0) ellipse (1 and 2);
        \draw[thin] (4,2) arc (90:-90:1 and 2);
        \draw[thin] (0,2) -- (4,2);
        \draw[thin] (0,-2) -- (4,-2);

        \draw (3,2) arc (90:30:1 and 2);
        \draw (1.5,-2) arc (-90:-30:1 and 2);

        \draw (0.86,-1) -- (2.36,-1) -- (3.86,1) -- (4.86,1);

        \draw (1.6,-.7) node[]{\(N_f\) \(\D5_L\)};
        \draw (4.3,.6) node[]{\(N_f\) \(\D5_R\)};

        \draw (3.2,1.4) node[]{\NS5'};
        \draw (1.8,-1.4) node[]{\NS5'};

        \draw (3.7,-.2) node[]{\((1,N_f)\)};
 
      \end{footnotesize}
    \end{scope}

  \end{tikzpicture}
  \caption{Geometric realization of the breaking of \(U(1)_A\)
    for compact \(x_3\). If the stacks of \D5s are moved apart to give a
    real mass, the \NS5'--brane cannot be closed in the direction \(x_3\) without a displacement in \(x_7\).}
  \label{U(1)A}
\end{figure}
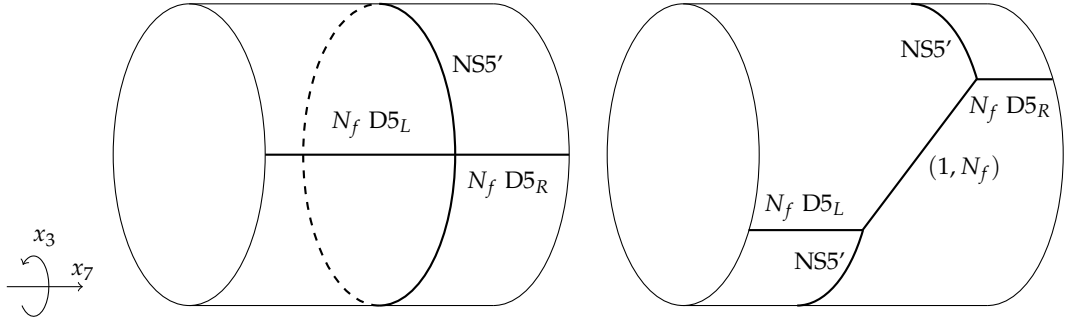
One might wonder what happens to the $U(1)_B$ and why it is still a good symmetry.
This symmetry is realized by the motion of the entire stack of \D5--branes with respect to the stack of \D3--branes
along \(x_3\). In this case the \D5--branes 
can slide together on the \NS{}--brane along $x_3$, without generating any $(1,N_f)$ fivebrane, and the symmetry is 
realized in the brane picture. 
Observe that in the three-dimensional picture this symmetry is gauged, it is 
associated to the motion of the \D3--branes rather than the motion of the \D5--branes.
\item The last symmetry is the topological $U(1)_J$ shifting the dual photon. In the \tIIB description
the dual photon is not visible and a geometric interpretation would require the lift of the configuration
to M--theory\footnote{Lifting our \tIIB picture to M--theory leads to a two-dimensional theory. While there are analogies with the three-dimensional systems at hand, we cannot use it to make precise statements. This is essential in our case as there are important physical effects that are related to global boundary conditions (\emph{i.e.} the fact that the \(x_3\) direction is periodic).} where the dual photon appears as a shift in $x_{10}$. This is related to the fact that this symmetry is not manifest in the Lagrangian but is an effect coming from the Bianchi identity. In the \tIIB description
we still have control of the real mass parameter associated to this symmetry, the \ac{fi} term.
In the brane picture it corresponds to the displacement of the \NS{} and \NS'--branes along $x_7$.
\end{itemize}

The same reduction can be performed in the magnetic picture. The discussion follows the one above and in the dual frame one obtains the 
same result when discussing the dual $U(N_f-N_c)$ gauge group.
Also in this case there are $N_f$ dual fundamentals and anti-fundamentals and a meson $M$.
The superpotential $M\,q \tilde q$ is combined with the $\eta'$ piece.
This is the geometrical version of the reduction of the four-dimensional Seiberg duality for \textsc{sqcd} 
in terms of T--duality and branes.
At finite radius, this theory can be treated as an effective three-dimensional theory if the
radius of the T--dual circle is large enough, and in this sense it represents a new three-dimensional duality.

\subsection{Aharony duality}
\label{sec:aharony-duality}

One can flow to the duality of Aharony by turning on 
real mass terms for some of the quarks~\cite{Aharony:2013dha}.
Here we reproduce this flow from the \tIIB brane perspective where the real masses are generated by moving \D5--branes in the $x_3$ direction.

\begin{figure}
  \begin{center}
    \begin{tikzpicture}
      \begin{scope}[shift={(-2,1)}]
        \begin{footnotesize}
          \draw[->] (0,3) -- (1,3);
          \draw[->] (.2,2.8) -- (.2,3.8);
          \draw (1,3.2) node[]{\(x_6\)};
          \draw (.2,4) node[]{\(x_3\)};
        \end{footnotesize}
      \end{scope}
      \begin{scope}[thick]
        \begin{footnotesize}

          \draw (0,1) -- (0,5); 
          \draw (-0.5,1) node[]{\NS5};
          \draw (0,3) -- (4,3); 
          \draw (1.5,3.5) node[]{\(N_c\) \D3}; 
          \draw[dashed] (4,1) -- (4,5);
          \draw (3.5,1) node[]{\NS5'};
          \draw[fill=white] (4,3) circle (5pt);
          \draw (4,3) node[cross=4pt]{};
          \draw (4.8,3) node[]{\(N_f\) \D5};
          \draw[fill=white] (4,1.5) circle (5pt);
          \draw (4,1.5) node[cross=4pt]{};
          \draw (4.6,1.5) node[]{\(1\) \D5};
          \draw[fill=white] (4,4.5) circle (5pt);
          \draw (4,4.5) node[cross=4pt]{};
          \draw (4.6,4.5) node[]{\(1\) \D5};
        \end{footnotesize}
      \end{scope}
      \begin{scope}[shift={(8,0)},thick]
        \begin{footnotesize}
          \draw (4,1) -- (4,5);
          \draw (3.5, 1) node[]{\NS5};
          \draw (0,3) -- (4,3);
          \draw (2.8, 3.5) node[]{\((N_f - N_c)\) \D3};
          \draw (0, 1.5) -- (4,1.5);
          \draw (3, 2) node[]{\(1\) \D3};
          \draw (0, 4.5) -- (4,4.5);
          \draw (3, 5) node[]{\(1\) \D3};
          \draw[dashed] (0,1) -- (0,5);
          \draw (-0.5,1) node[]{\NS5'};
          \draw[fill=white] (0,3) circle (5pt);
          \draw (0,3) node[cross=4pt]{};
          \draw (-0.8,3) node[]{\(N_f\) \D5};
          \draw[fill=white] (0,1.5) circle (5pt);
          \draw (0,1.5) node[cross=4pt]{};
          \draw (-0.7,1.5) node[]{\(1\) \D5};
          \draw[fill=white] (0,4.5) circle (5pt);
          \draw (0,4.5) node[cross=4pt]{};
          \draw (-0.7,4.5) node[]{\(1\) \D5};
        \end{footnotesize}

      \end{scope}
    \end{tikzpicture}
  \end{center}
\caption{Electric and magnetic brane system for \textsc{sqcd} on $\setR^3\times S^1$}
\label{fig:electric-magnetic-SQCD}
\end{figure}
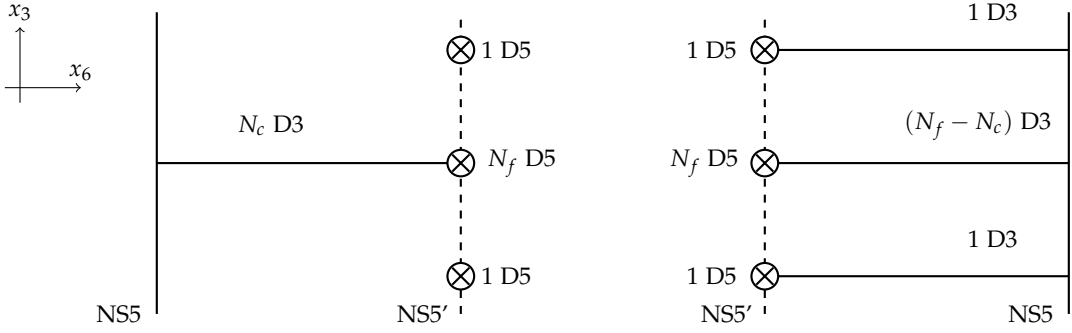

Consider the case with $N_f+2$ flavors. The flow is generated by breaking the  
$SU(N_f+2)^2$ flavor symmetry down to $SU(N_f)^2 \times U(1)_A$. 
We study the theory at an energy scale $E < 1/ \widetilde R_3$ (where $\widetilde R_3 = \alpha'/R_3$ is the T--dual radius)
so that 
$x_3$ is effectively non-compact. In this large mass limit, the $\eta$--superpotential disappears and the axial $U(1)_A$ is restored. 
The flavor symmetry is broken in the brane description by moving 
one \D5--brane in the $x_3>0$ direction and one \D5 in the $x_3<0$ direction (the stack of $N_c$ \D3--branes sits at $x_3=0$).
The configuration that we obtain is 
shown on the left-hand side of Figure~\ref{fig:electric-magnetic-SQCD}. 
We end up with a $U(N_c)$ gauge theory with 
$N_f$ fundamentals an anti-fundamentals and vanishing superpotential. This is the electric theory studied in 
Aharony.

The dual picture is shown on the right-hand side of Figure~\ref{fig:electric-magnetic-SQCD}.
In this case, each \D5--brane drags one \D3 along the $x_3$ direction and there are three gauge sectors, $U(N_f-N_c) \times U(1)^2$ as expected from field theory.
The chiral multiplets connecting the two sectors acquire a large mass as long as the
\D5--branes separate along $x_3$. However, in this case there are one fundamental and one anti-fundamental massless flavors and a meson with a superpotential interaction in each $U(1)$ sector.
The three sectors interact on the Coulomb branch. Indeed,
if we consider the Coulomb branch in the brane picture, the \D3s separate along $x_3$ at equal distance.
The $U(1)$ sectors are crucial for understanding the structure of the moduli space of the dual phase.
In the large mass limit, when the two \D5s are far away in the $x_3$ direction, two \D3s of the stack of
$N_f-N_c$ \D3--branes are free to move, and this parametrizes the Coulomb branch.
Differently from the electric case, these \D3s cannot be pushed to $\pm \infty$, because there
are two extra \D3--branes, the \(U(1)\) sectors.
This constraint is reflected in an interaction between the $U(N_f-N_c)$ sector and the $U(1)$ sectors, this interaction is an \ac{ahw} 
superpotential, coming from the broken $U(N_f-N_c+2)$ theory.
At the level of the brane system it is due to the \D1--strings 
between the \NS{}--branes, the two \D3s that are pushed far away in the $x_3$ direction by 
the \D5s, and the two \D3s that parametrize the Coulomb branch of the $U(N_f-N_c)$ sector.

As discussed in Section~\ref{sec:superpotential-from-branes} the  
\D1--branes create the non-perturbative \ac{ahw} superpotential
\begin{equation}
\label{eq:monmox}
W_{\text{AHW}} = e^{\Sigma^{(1)}-\Sigma_{\widetilde N_c}}+ e^{\Sigma_1- \Sigma^{(2)}}+ \eta' e^{\Sigma^{(2)}-\Sigma^{(1)}}
=
\tilde y_1 y + \tilde y y_2 + \eta'  \tilde y_2 y_1.
\end{equation}
Here $\Sigma^1$ ($\Sigma^2$) corresponds to the Coulomb branch coordinate of the $1$st ($2$nd) $U(1)-$sector.
Similarly, $\Sigma_1$ and $\Sigma_{\widetilde N_c}$ correspond to the Coulomb branch coordinates that remain unlifted by $\eqref{WD1}$, 
which we do not repeat here.

By going to large $R_3$, we can think of the dual version of each $U(1)$ sector
at $x_3 > 0 $ and $x_3 < 0$.
Recall that a $U(1)$ gauge theory with one pair of fundamental and anti fundamental  is dual to 
the XYZ model.
This duality is realized in the brane system by exchanging the \NS{} and the \D5--branes.
Our $U(1)$ sectors are similar to the gauge theory dual of the XYZ models, yet there is a slight modification due to the additional superpotential interaction $M^{(i)} q^{(i)} \tilde q^{(i)}$ 
in each sector $i = 1,2$.
By duality, the $i$th sector corresponds  to a deformation of the XYZ model, where the singlets $X, Y$ and $Z$ are identified with the meson $N^{(i)}= q^{(i)} \tilde q^{(i)}$
and the Coulomb branch coordinates $y_i,\tilde y_i$.
The resulting superpotential  is given by
\begin{equation}
  W_{\text{m}}^{(i)} = M^{(i)} N^{(i)} + N^{(i)} y_i \tilde y_i .
\end{equation}
Upon integrating out the massive fields, 
the superpotential $W_{\text{m}}^{(1)} +  W_{\text{m}}^{(2)} + W_{\text{AHW}} $
becomes the interaction $y Y + \tilde y \widetilde Y$ expected for the Aharony duality.
This is done by interpreting 
$\tilde y_1$ and $y_2$  as the singlets $Y$ and $\widetilde Y$ corresponding to the monopole operators of the electric phase. 

We conclude this section with a general remark.
For four-dimensional Seiberg duality, the brane construction is well understood. Using dimensional reduction, we have been able to describe Aharony duality in terms of branes, since the four-dimensional theories naturally provide a \textsc{uv} completion.  This idea can be generalized and applied to other \textsc{uv} 
pairs flowing to Aharony duality. 
One can for example use the \textsc{rg} flow from \ac{gk} duality to Aharony duality recently studied in~\cite{Intriligator:2013lca,Khan:2013bba,Amariti:2013qea,Amariti:2014lla}.

\section{Extensions}
\label{sec:extensions-outlook}

In this section we introduce some extensions of the procedure explained above. 
The dimensional reduction of four-dimensional dualities to three dimensions can 
be generalized to more general gauge theories~\cite{Aharony:2013dha}.
Many details of this construction are model-dependent: gauge group and matter content of a theory determine the \textsc{ir} dynamics which in turn affect
the structure of the reduced theory.
The brane construction provides a unified realization of the reduction of four-dimensional dualities
and here we apply it to some dualities with a \tIIA description.

We first use the brane picture to flow from the duality of Aharony to \ac{gk} duality. This flow generates \ac{cs} levels and is of importance for the 
analysis of other dualities in three dimensions.

Next we will discuss three-dimensional dualities for quiver gauge theories.
By generating \ac{cs} levels we obtain quivers describing the \textsc{ir} dynamics of \M2--branes.
This is of interest as it can link the dynamics of \M2--branes to AdS$_5$/CFT$_4$.

Another generalization includes theories with a richer matter content.
Here we explain how to reduce the dualities of~\cite{Kutasov:1995ss} for theories with adjoint matter using the brane description.
Our analysis reproduces the field theory results of~\cite{Nii:2014jsa}.

\subsection{Giveon--Kutasov duality}
\label{sec:GK-duality}

In this subsection we study the \textsc{rg} flow of four-dimensional Seiberg duality 
to \ac{gk} duality for three-dimensional $U(N)_k$ \textsc{sqcd},
where $k$ is the Chern--Simons level.
In field theory this has been studied in~\cite{Willett:2011gp}. One can start from the Aharony 
electric phase with $N_f+k$ flavors and use the global symmetry to assign a large 
mass to $k$ fundamentals and anti-fundamentals, with the same (e.g. positive) sign.
At large mass it generates the level $k$ in the 
electric theory. In the dual theory after integrating out the massive fields (also the 
monopoles acquire a mass term) one is left with $U(N_f-N_c+|k|)_{-k}$ \textsc{sqcd} with 
superpotential $W_{\text{m}}=M\, q \tilde q$. 
In the brane picture we first consider $N_f+k$ \D5--branes instead of $N_f$.
Then we separate $k$ \D5s on the \NS'--brane, and obtain two semi-infinite stacks.
We move half of the stack to $x_3>0$ and the other half to $x_3<0$.
This process creates a $(1,k)$ five-brane.
In the limit when the $(1,k)$ fivebrane becomes infinite a \ac{cs} term 
is generated~\cite{Kitao:1998mf,Aharony:1997ju}.
At the level of field theory this reproduces the electric side of the \ac{gk} duality.
In the dual phase the situation is more complicated. The number of \D3s created 
by this process is still $N_f+k+2$, the gauge theory is now $U(1)^2 \times U(N_f-N_c+k)$.
The motion of the \D5s along $x_3$ generates a non-trivial \ac{fi} term for the $U(1)$ sectors. This term is proportional 
to the axial real mass (in the flow $U(1)_J$ and $U(1)_A$ do mix indeed). This implies that the 
monopoles are massive as expected and they disappear.
At the end one is left with the dual \ac{gk} configuration.

\subsection{Product groups}

Consider a class of four-dimensional gauge theories 
that consist of products of $G$ $U(N)$ gauge groups with bifundamental and fundamental
matter fields. This construction was first discussed in four dimensions in~\cite{Uranga:1998vf}
for the case of the conifold.
These systems have many possible Seiberg-dual phases in four dimensions.
Here we discuss how this duality reduces to three dimensions along the lines explained above.
We engineer this construction in a brane system and eventually relate this duality to the
toric duality for \M2--branes discussed in~\cite{Amariti:2009rb}.

\begin{figure}
  \begin{center}
    \begin{tikzpicture}
      \node[inner sep=0pt] at (0,0) {\includegraphics[width=.45\textwidth]{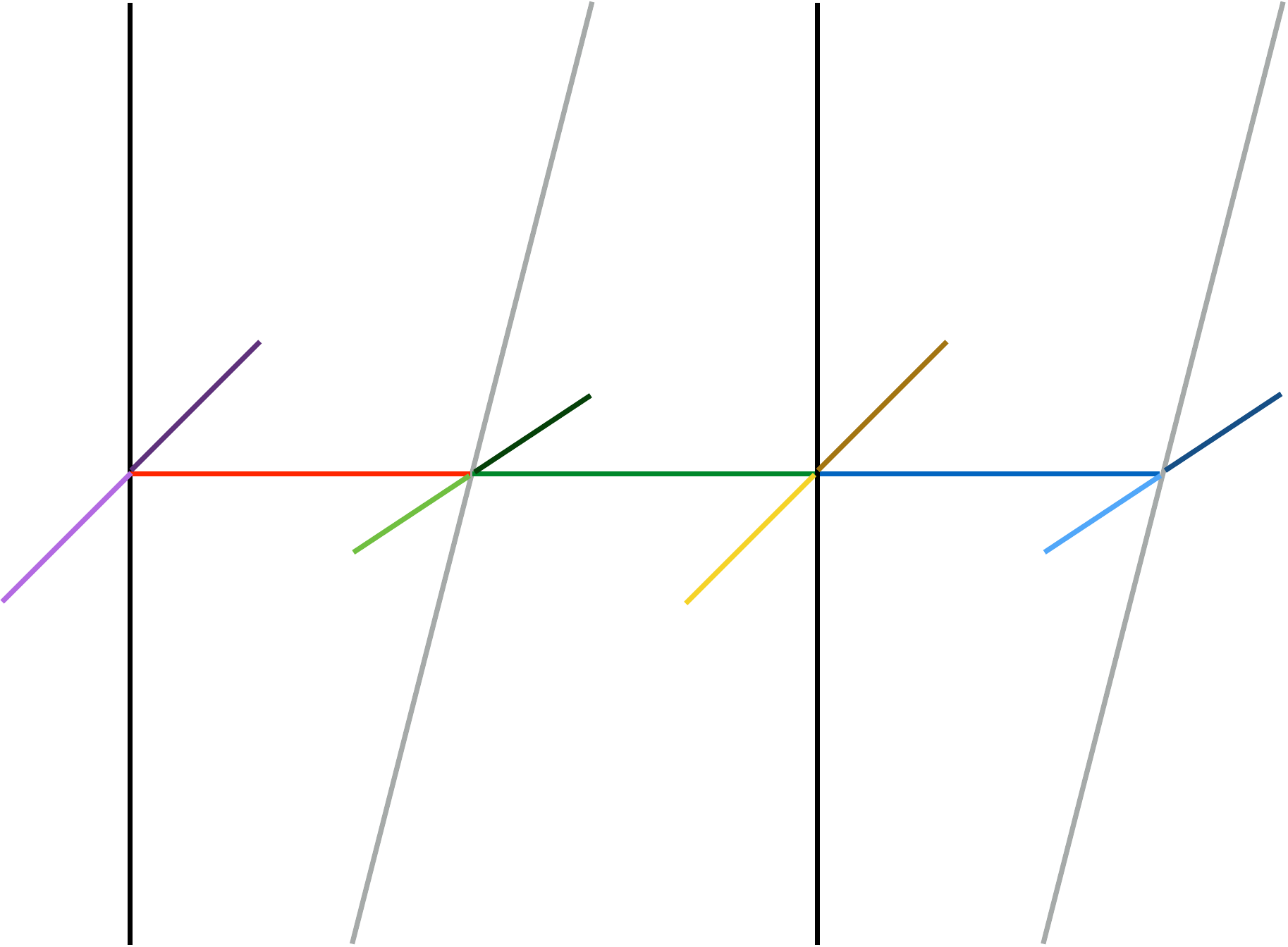}};
      \node[inner sep=0pt] at (7.2,0) {\includegraphics[width=.45\textwidth]{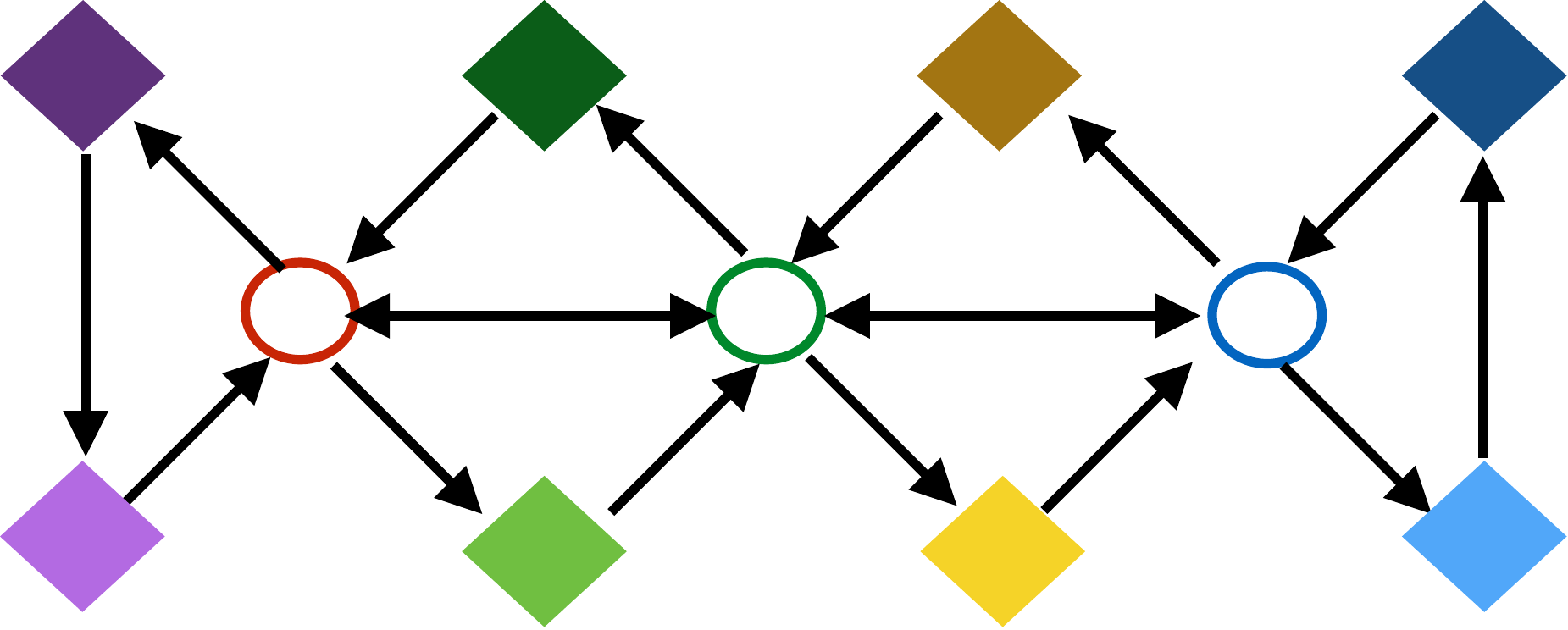}};
      \begin{scope}[shift={(-4,-2.7)},every node/.style={anchor=base west}]
        \begin{footnotesize}
          \draw (1,0) node[]{\NS5};
          \draw (2,0) node[]{\NS5'};
          \draw (4.5,0) node[]{\NS5};
          \draw (5.5,0) node[]{\NS5'};

          \draw (2,3.5) node[]{\D6};
          \draw (3.7,3.2) node[]{\D6'};
          \draw (5.5,3.5) node[]{\D6};
          \draw (7.1,3.2) node[]{\D6'};

          \draw (1.8,2.3) node[]{\D4};
          \draw (3.5,2.3) node[]{\D4};
          \draw (5.2,2.3) node[]{\D4};

          \draw (9.1,2.1) node[]{1};
          \draw (11,2.1) node[]{2};
          \draw (12.9,2.1) node[]{3};
          
        \end{footnotesize}
      \end{scope}
    \end{tikzpicture}
  \end{center}
  \caption{Brane description and linear quiver representing 
    the product of four-dimensional $SU(N)$ gauge group with flavor.
    We use the colors to identify the gauge and flavor symmetries associated to the 
    \D3 and \D5--branes respectively.}
  \label{fig:brane-linear-quiver}
\end{figure}
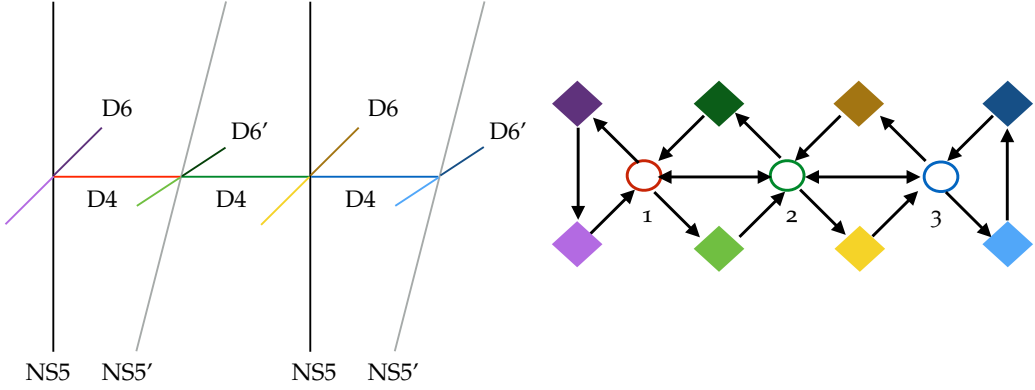

We start by considering  a system of \D4--branes suspended between $G+1$ \NS{}~and \NS'--branes. On each \NS{} (\NS') brane we put a \D6 (\D6') brane.
The \NS{}--branes are extended along \(0123(45,89)_\alpha\), where \((x,y)_\alpha\) denotes a rotation in the directions \(x,y\). The \NS' branes are extended along \(0123(45,89)_{\alpha'}\), where the prime denotes the \textsc{susy}-preserving complementary angle. Furthermore, \D6 branes are along \(0123(45,89)_{\alpha}7\) and \D6' along \(0123(45,89)_{\alpha'}7\)%
. In the
language of quiver gauge theory we have a product of $SU(N_i)$ gauge groups, where each $N_i$ 
is the number of \D4--branes between two consecutive fivebranes.
There are bifundamental fields $Q_{ij}$ connecting two consecutive gauge nodes. In our notation, \(Q_{ij}\) is in the fundamental representation of $SU(N_i)$ and in the anti-fundamental representation of $SU(N_j)$.
There are also fundamental matter fields, associated to the \D6 and the \D6'--branes:  each gauge factor $U(N_i)$ has a pair of fundamental and anti-fundamental
$(q_i,\tilde q_i)$ coming from the  \D6 (or \D6') on the right and a pair $(p_i,\tilde p_i)$ from the one on the left. 
The number of \D6s and \D6's has to be
chosen consistently with the s--rule~\cite{Hanany:1996ie,Kitao:1998mf}. 
In general there are $SU(N_f)$ and $U(1)$ flavor groups, $G$ $U(1)$ baryonic symmetry groups and the $U(1)$ $R$--symmetry groups, but some of the $U(1)$ flavor symmetries are anomalous.
The brane system and the quiver for this case are shown in Figure~\ref{fig:brane-linear-quiver}.

There are two types of superpotential interactions. One involves only the bifundamental 
fields\footnote{We describe in detail the case with alternating \NS{} and \NS'--branes but more general configurations are possible.}
\begin{equation}
  W_{\text{bif}}^{(i)} = (-1)^i Q_{i-1,i} Q_{i,i+1} Q_{i+1,i} Q_{i,i-1} ,
\end{equation}
where the sign comes from the alternate signs of the adjoint masses.
The interactions of the fundamental fields are
\begin{equation}
  W_{\text{f}}^{(i)} = p_i Q_{i,i+1} \tilde p_i + \tilde q_i Q_{i+1,i} q_i
\end{equation}
and
\begin{align} 
  \label{eq:W-linear-quiver}
  W_{\text{l}} = p_1 M \tilde q_1 + q_G N \tilde p_G,
\end{align} 
where $M$ and $N$ are gauge singlets that transform in the bifundamental representation of the first and the last pairs of flavor groups respectively.

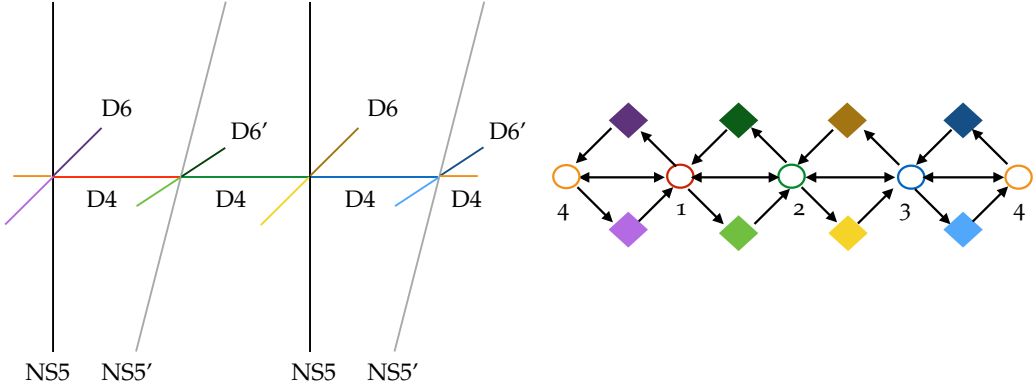
\begin{figure}
  \begin{center}
    \begin{tikzpicture}
      \node[inner sep=0pt] at (0,0) {\includegraphics[width=.45\textwidth]{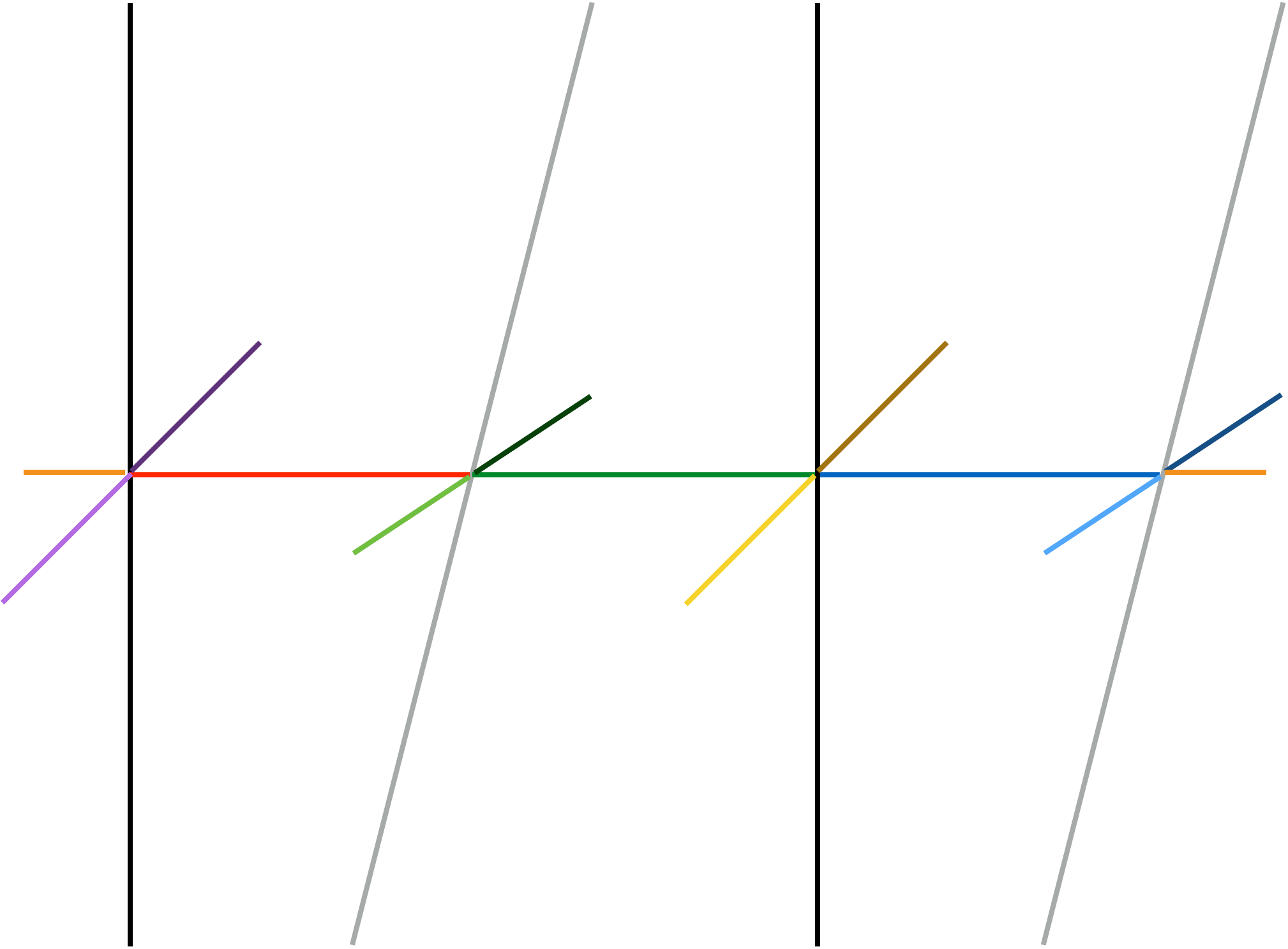}};
      \node[inner sep=0pt] at (7.2,0) {\includegraphics[width=.45\textwidth]{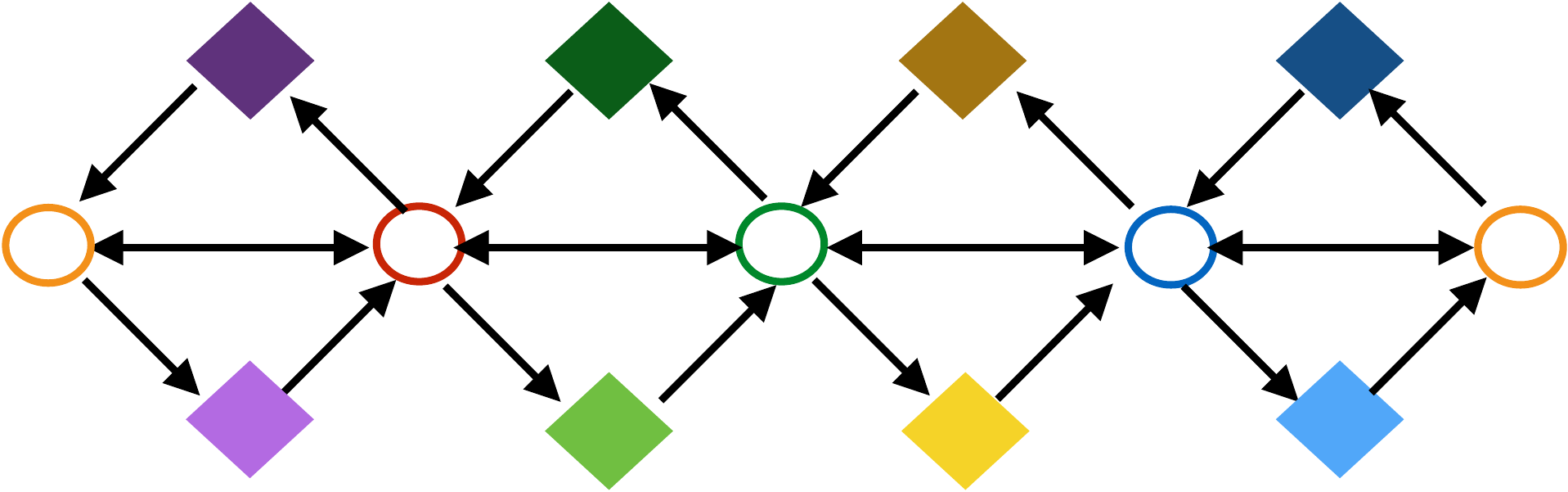}};
      \begin{scope}[shift={(-4,-2.7)},every node/.style={anchor=base west}]
        \begin{footnotesize}
          \draw (1,0) node[]{\NS5};
          \draw (2,0) node[]{\NS5'};
          \draw (4.5,0) node[]{\NS5};
          \draw (5.5,0) node[]{\NS5'};

          \draw (2,3.5) node[]{\D6};
          \draw (3.7,3.2) node[]{\D6'};
          \draw (5.5,3.5) node[]{\D6};
          \draw (7.1,3.2) node[]{\D6'};

          \draw (1.8,2.3) node[]{\D4};
          \draw (3.5,2.3) node[]{\D4};
          \draw (5.2,2.3) node[]{\D4};
          \draw (6.6,2.3) node[]{\D4};

          \draw (8,2.2) node[]{4};
          \draw (9.55,2.2) node[]{1};
          \draw (11.1,2.2) node[]{2};
          \draw (12.5,2.2) node[]{3};
          \draw (14,2.2) node[]{4};

        \end{footnotesize}
      \end{scope}
    \end{tikzpicture}
  \end{center}
  \caption{Brane description and circular quiver representing 
    the product of four-dimensional $SU(N)$ gauge group with flavor.
    We use the colors to identify the gauge and flavor symmetries associated to the 
    \D3 and \D5--branes respectively.}
  \label{fig:brane-circular-quiver}
\end{figure}

Up to now we have considered \(x_6\) to be infinite and the \D4--branes to be bounded by \NS5--branes. 
We can allow a slightly different situation, where the direction $x_6$ is compact and the quiver is circular. 
This case is shown in Figure~\ref{fig:brane-circular-quiver}. 
In the circular case the total number of gauge groups is $G+1$ and the superpotential $\eqref{eq:W-linear-quiver}$ vanishes.
These systems have a large number of Seiberg-dual phases in four dimensions, which may in principle be reduced to $3$d.

\bigskip

At this point we can compactify the direction $x_3$ and perform the by now familiar reduction.
Observe that in the three-dimensional case we consider the baryonic symmetry as gauged, and the gauge groups are enhanced to $U(N_i)$. First we discuss the field theory reduction and then we show how this mechanism is realized in the brane setup.
We first study the case of linear quivers and then switch to the circular ones.

\paragraph{Field theory reduction.}

By putting the theory on $\setR^3 \times S^1$, the additional superpotential
\begin{equation}
  W_\eta^{(i)}  = \eta_i Y_i \widetilde Y_i,
\end{equation}
is generated, where $Y_i$ refers to the monopole of the $i$-th gauge group with magnetic flux $(1,0,\dots,0)$.
This superpotential breaks the flavor symmetries that are anomalous in $4$d.
By proceeding similarly on the magnetic side we obtain a set of equivalent phases which generalize the dualities with $\eta$--superpotential proposed in~\cite{Aharony:2013dha}.

\begin{itemize}
\item As in the case with one gauge group, we
  can flow to an Aharony-like duality for these quivers. We assign
  large and opposite real masses to the same amount of fundamentals and anti-fundamentals in each
  flavor sector. This procedure eliminates the $\eta$--superpotential
  and restores the $4$d--anomalous flavor symmetries.
  In the dual theory the real masses are given consistently
  with the global symmetries; in addition one has to choose a
  non-trivial vacuum. This eliminates also the magnetic $\eta'$--superpotential 
  and higgses the dual gauge group. The higgsing
  generates \ac{ahw} superpotentials as in Eq.~$\eqref{eq:W-preAharony}$.
  This leads to the Aharony duality for this class of quiver gauge
  theories.
\item Finally, by integrating out the remaining fundamental fields one can
  generate \ac{cs} levels $(k+i)$ for the $i = 1,\dots,G$ gauge groups.
  This construction generalizes the \ac{gk} duality~\cite{Aharony:2008gk,Amariti:2009rb}.
\item For the circular quiver the $i$th \D5--brane contributes with an opposite factor to the \ac{cs}
of the $(i-1)$st and of the $i$th gauge group. This implements the constraint $\sum k_i = 0$
and we obtain the models studied in~\cite{Amariti:2009rb}, 
the generalization to three dimensions of the $L^{aba}$
theories~\cite{Benvenuti:2005ja,Franco:2005sm,Butti:2005sw}.
They represent the moduli space of a stack of \M2--branes probing a $CY_{4}$ singularity.
It has been shown in~\cite{Amariti:2009rb} that in these theories the three-dimensional Seiberg duality (\emph{i.e.} the exchange of two consecutive five-branes) is the same as toric duality.
\end{itemize}

\paragraph{Brane picture.}

In this section we discuss the derivation of the 
dualities discussed above as seen from the brane picture.

We start with a configuration of \D4--branes
suspended between \NS5 and \NS5'--branes.
For simplicity we choose the case with $N_c$ \D4--branes 
between each pair of five-branes even if more general configurations are possible.
Then we add the \D6 and \D6'--branes on the \NS5 and \NS5'--branes. Finally we compactify the $x_3$ direction and T--dualize,  generalizing the construction of~\cite{Cremonesi:2010ae}.

\begin{itemize}
\item \D1--branes wrapping the circle in $x^3$ create an
  $\eta$--superpotential for each gauge group. When doing the
  analogous operation in the magnetic phase, this produces the brane
  picture of the duality with $\eta'$--superpotential for the linear and circular quivers.
\item Upon sending pairs of \D5--branes to large distances on $x^3$, the
  \D1--strings disappear and we reproduce the Aharony duality for
  quivers as in Section~\ref{sec:aharony-duality}.
\item As discussed in Section~\ref{sec:GK-duality}, we can
  generate \ac{cs} levels from the brane picture and generalize the
  \ac{gk} duality.
\item In the circular case, if we integrate out all the flavors, the models correspond to the generalized $L^{aba}_{\{k_i\}}$ \tIIB description of \M2--branes probing CY$_4$ singularities. 
These theories have been shown to enjoy a toric duality in three dimensions~\cite{Amariti:2009rb}.
This connects this three-dimensional toric duality 
with four-dimensional Seiberg duality. 
\item Another class of models describing \M2--branes probing CY$_4$ singularities consists of circular vector-like quivers with bifundamental matter and chiral flavor. These models have been proposed in~\cite{Benini:2009qs,}.
They can be obtained by assigning a large mass only to some of the flavors. This generates (semi-)~integer 
\ac{cs} levels and chiral matter. In the brane picture this is done by sending only half (\emph{i.e.} $N_f$) of the \D5s which are broken on the \NS{}--branes to large distances. The \ac{cs}~terms are due to the semi-infinite $(1,N_f)$ fivebranes generated by this procedure. 
\end{itemize}

We conclude this section but mentioning the possibility of reducing four dimensional chiral quiver gauge theories.
Here the situation is more complicate, because their three dimensional
counterpart, when decorated with \ac{cs} levels do not have a simple
interpretation in terms of \M2--branes (for example the free energy
does not scale as $N^{3/2}$~\cite{Jafferis:2011zi} as expected from the conjectured gravity dual\footnote{See~\cite{Amariti:2011uw,Gulotta:2011aa} for a possible solution to this problem.}).
It would be interesting to study how the dimensional reduction of these
theories can shed light
on this problem.

\subsection{Adjoint matter}

In this section we study the dimensional reduction of four-dimensional 
Seiberg duality with adjoint matter to three dimensions.
From the field theory point of view, this has been first studied in~\cite{Nii:2014jsa,Amariti:2014iza}.
Here we first review the field theoretical construction and then  we discuss the 
mechanism from the brane perspective.
In four dimensions this is called \ac{kss} duality.

The electric model is an $SU(N_c)$ gauge theory with $N_f$ fundamentals $Q$ and $N_f$ 
anti-fundamentals
$\widetilde Q$  and one adjoint field $X$. There is a superpotential coupling 
\begin{equation}
\label{WXk}
W_{\text{el}}^{\text{KSS}} = \Tr X^{n+1}
\end{equation}
with $n \leq N_c$. When $n N_f > N_c$ there  is a magnetic  description.
The dual field theory has an $SU(N_c)$ gauge group with $N_f$ dual quarks
 $q$ and $\tilde q$ and an adjoint $Y$.
There are also electric mesons $M_j$ appearing as elementary degrees of freedom in this
dual phase. They have the form
\begin{equation}
M_j = Q X^{j} \widetilde Q \quad \quad j=0,\dots,n-1 .
\end{equation}
The superpotential of the dual theory is given by
\begin{equation}
W_{\text{m}}^{\text{KSS}} = \Tr Y^{n+1} + \sum_{j=0}^{n-1} M_j q Y^{n-j-1} \tilde q .
\end{equation}
The global symmetry group is $SU(N_f)^2 \times U(1)_B \times U(1)_R$.
There is also an axial symmetry that is anomalous in four dimensions.

The three-dimensional reduction of this duality has been discussed in~\cite{Nii:2014jsa}, generalizing the procedure of~\cite{Aharony:2013dha}.
Here we sketch the main steps of this reduction.
One first compactifies the theory on a circle of finite radius $R_3$. Then one realizes that in this case the 
analogue of the $\eta$--superpotential is generated in both the electric and the magnetic phases.
These superpotentials are of the form 
\begin{align}
\label{etaadj}
W_{\eta} &= \eta \sum_{j=0}^{n-1} T_{j} \widetilde T_{n-1-j} ,
&
W_{\eta'} &= \eta' \sum_{j=0}^{n-1} t_{j} \tilde t_{n-1-j} ,
\end{align}
where $T_j$ and $\widetilde T_j$ are monopole and anti-monopole operators.
They can be written in terms of the original monopoles $Y$ as 
$T_j=Y X^{j}$. 
By using the same arguments of the \textsc{sqcd} case, the electric and the magnetic theories obtained by considering the superpotentials Eq.~(\ref{etaadj}) are Seiberg--dual in three dimensions.

The next step consists in flowing from these dualities to the Aharony--like case. This last
duality has been first introduced in three dimensions by \ac{kp}.
This flow is similar to the one discussed for ordinary \textsc{sqcd}. 
One can first add a superpotential of the form
\begin{equation}
\label{polyX}
W_{\text{el}}^{P(X)} = \sum_{j=1}^{n} \alpha_j  X^j  
\end{equation}
in the adjoint fields, and analogously on the magnetic side.
This superpotential breaks the gauge group into a product of decoupled \textsc{sqcd}s, 
each with gauge symmetry  $U(r_i)$ with $\sum_i r_i = N_c$, $N_f$ fundamentals
and anti-fundamentals and no adjoint. 
In each sector the reduction works like in \textsc{sqcd}. One can finally send the $\alpha_i$ 
couplings to zero. As shown in~\cite{Amariti:2014iza}, this procedure is consistent
and can be applied also to the magnetic theory.
Eventually the \ac{kp} duality is obtained.
The charges of the fields under the global symmetries in the \ac{kp} duality are shown in Table~\ref{tab:charges-KP}.
\begin{table} 
\centering
\begin{tabular}{cccccc}
  \toprule
                  & \( SU(N_f)_L      \) & \( SU(N_f)_R      \) & \( U(1)_A \) & \( U(1)_R                               \) & \( U(1)_J \)\\
\midrule
\(Q                \) & \( N_f            \) & \( 1              \) & \( 1      \) & \( \Delta                               \) & \( 0 \)     \\
\(\widetilde Q     \) & \( 1              \) & \( \overline{N_f} \) & \( 1      \) & \( \Delta                               \) & \( 0   \)   \\
\(X                \) & \( 1              \) & \( 1              \) & \( 0      \) & \( \frac{2}{n+1}                        \) & \( 0 \)     \\
\midrule
\(q                \) & \( \overline{N_f} \) & \( 1              \) & \( -1     \) & \( \frac{2}{n+1}-\Delta                 \) & \( 0 \)     \\
\(\tilde q     \) & \( 1              \) & \( N_f            \) & \( - 1    \) & \( \frac{2}{n+1}-\Delta                 \) & \( 0 \)     \\
\(Y                \) & \( 1              \) & \( 1              \) & \( 0      \) & \( \frac{2}{n+1}                        \) & \( 0 \)     \\
\(M_j              \) & \( N_f            \) & \( \overline{N_f} \) & \( 2      \) & \( 2 \Delta+\frac{2j}{n+1}              \) & \( 0  \)    \\
\(T_{j}            \) & \( 1              \) & \( 1              \) & \( -N_f   \) & \( N_f(1-\Delta)-\frac{2}{n+1}(N_c-1-j) \) & \( 1  \)    \\
\(\widetilde T_{j} \) & \( 1              \) & \( 1              \) & \( -N_f   \) & \( N_f(1-\Delta)-\frac{2}{n+1}(N_c-1-j) \) & \( - 1 \)  \\
\(t_j              \) & \( 1              \) & \( 1              \) & \( N_f    \) & \( N_f(\Delta-1)+\frac{2}{n+1}(N_c+1+j) \) & \( -1\)    \\ 
\(\tilde{t}_j  \) & \( 1              \) & \( 1              \) & \( N_f    \) & \( N_f(\Delta-1)+\frac{2}{n+1}(N_c+1+j) \) & \( 1 \) \\
\bottomrule
\end{tabular}
  \caption{Global charges for the configuration in the \ac{kp} duality}
  \label{tab:charges-KP}
\end{table}

Observe that one could break the gauge theory into a decoupled set of \textsc{sqcd}s
also in the four-dimensional case and perform the reduction in the broken case. 
This would have modified Eq.~(\ref{etaadj}) to
\begin{align}
\label{brokenetaadj}
  W_{\eta} &= \sum_{i=1}^{n} \eta_i Y_{i} \widetilde Y_{i} ,
&
  W_{\eta'} &=  \sum_{i=1}^{n} \eta_i' y_{i} \tilde y_{i} ,
\end{align}
where the subscript $i$ labels the $U(r_i)$ gauge group and \(y_i\), \(Y_i\) are the monopoles in each phase.
The relation between the superpotential in Eq.~\eqref{etaadj} and the
one in Eq.~(\ref{brokenetaadj})
is obtained by the scale matching relation~\cite{Kutasov:1995ss}
\begin{equation}
\Lambda^{2N_c-N_f} = \Lambda^{3r_i - N_f} \prod_{j\neq i} (\omega_i - \omega_j)^{r_i-2r_j},
\end{equation}
where $\omega_i$ parametrize the \ac{vev} of the adjoint fields
breaking $U(N_c)$ to 
$\prod U(r_i)$. An analogous relation can be written for the magnetic case.
This last observation is useful for the reduction of the \ac{kss} duality from the brane 
picture.

\begin{figure}
  \begin{center}
    \begin{tikzpicture}
      \begin{scope}[shift={(-2,-3.5)}]
        \begin{footnotesize}
          \draw[->] (0,3) -- (1,3);
          \draw[->] (.2,2.8) -- (.2,3.8);
          \draw (1,3.2) node[]{\(x_6\)};
          \draw (.2,4) node[]{\(x_{4,5}\)};
        \end{footnotesize}
      \end{scope}

      \begin{scope}[thick]
        \begin{footnotesize}
          \draw (0,0) -- (0,8);
          \draw (-0.5,0) node[]{\NS5'};

          \draw (0,7) -- (2.5,7);
          \draw (1.5,6.5) node[]{\(r_1\) \D3};
          \draw[fill=black] (2.5,7) node{} circle (5pt);
          \draw (2.5,7.5) node[]{\NS5};
          \draw[fill=white] (0,7) circle (5pt);
          \draw (0,7) node[cross=4pt]{};
          \draw (-0.75,7) node[]{\(N_f\) \D5};

          \draw (0,5) -- (2.5,5);
          \draw (1.5,4.5) node[]{\(r_2\) \D3};
          \draw[fill=black] (2.5,5) node{} circle (5pt);
          \draw (2.5,5.5) node[]{\NS5};
          \draw[fill=white] (0,5) circle (5pt);
          \draw (0,5) node[cross=4pt]{};
          \draw (-0.75,5) node[]{\(N_f\) \D5};

          \draw (0,3) -- (2.5,3);
          \draw (1.5,2.5) node[]{\dots};
          \draw[fill=black] (2.5,3) node{} circle (5pt);
          \draw (2.5,3.5) node[]{\NS5};
          \draw[fill=white] (0,3) circle (5pt);
          \draw (0,3) node[cross=4pt]{};
          \draw (-0.75,3) node[]{\(N_f\) \D5};

          \draw (0,1) -- (2.5,1);
          \draw (1.5,0.5) node[]{\(r_n\) \D3};
          \draw[fill=black] (2.5,1) node{} circle (5pt);
          \draw (2.5,1.5) node[]{\NS5};
          \draw[fill=white] (0,1) circle (5pt);
          \draw (0,1) node[cross=4pt]{};
          \draw (-0.75,1) node[]{\(N_f\) \D5};

          \draw (1,-.5) node[]{(a)};

        \end{footnotesize}
      \end{scope}

      \begin{scope}[shift={(5,0)},thick]
        \begin{footnotesize}
          \draw (1.5,0) -- (1.5,8);
          \draw (1,0) node[]{\NS5'};

          \draw (0,7.05) -- (2.5,7.05);
          \draw (0, 6.95) -- (1.5,6.95);
          \draw (.5,6.5) node[]{\((N_f - r_1)\) \D3};
          \draw[fill=black] (2.5,7) node{} circle (5pt);
          \draw (2.5,7.5) node[]{\NS5};
          \draw[fill=white] (0,7) circle (5pt);
          \draw (0,7) node[cross=4pt]{};
          \draw (-0.75,7) node[]{\(N_f\) \D5};

          \draw (0,5.05) -- (2.5,5.05);
          \draw (0,4.95) -- (1.5,4.95);
          \draw (.5,4.5) node[]{\((N_f - r_2)\) \D3};
          \draw[fill=black] (2.5,5) node{} circle (5pt);
          \draw (2.5,5.5) node[]{\NS5};
          \draw[fill=white] (0,5) circle (5pt);
          \draw (0,5) node[cross=4pt]{};
          \draw (-0.75,5) node[]{\(N_f\) \D5};

          \draw (0,3.05) -- (2.5,3.05);
          \draw (0,2.95) -- (1.5,2.95);
          \draw (.5,2.5) node[]{\dots};
          \draw[fill=black] (2.5,3) node{} circle (5pt);
          \draw (2.5,3.5) node[]{\NS5};
          \draw[fill=white] (0,3) circle (5pt);
          \draw (0,3) node[cross=4pt]{};
          \draw (-0.75,3) node[]{\(N_f\) \D5};

          \draw (0,1.05) -- (2.5,1.05);
          \draw (0,0.95) -- (1.5,0.95);
          \draw (.5,0.5) node[]{\((N_f - r_n)\) \D3};
          \draw[fill=black] (2.5,1) node{} circle (5pt);
          \draw (2.5,1.5) node[]{\NS5};
          \draw[fill=white] (0,1) circle (5pt);
          \draw (0,1) node[cross=4pt]{};
          \draw (-0.75,1) node[]{\(N_f\) \D5};

          \draw (1,-.5) node[]{(b)};

        \end{footnotesize}
      \end{scope}

      \begin{scope}[shift={(9,0)},thick]
        \begin{footnotesize}
          \draw (2.5,0) -- (2.5,8);
          \draw (2,0) node[]{\NS5'};

          \draw (0,7) -- (2.5,7);
          \draw (1.2,6.5) node[]{\((N_f - r_1)\) \D3};
          \draw[fill=black] (0,7) node{} circle (5pt);
          \draw (1.9,7.5) node[]{\(N_f\) \D5};
          \draw[fill=white] (2.5,7) circle (5pt);
          \draw (2.5,7) node[cross=4pt]{};
          \draw (0,7.5) node[]{\NS5};

          \draw (0,5) -- (2.5,5);
          \draw (1.2,4.5) node[]{\((N_f - r_2)\) \D3};
          \draw[fill=black] (0,5) node{} circle (5pt);
          \draw (1.9,5.5) node[]{\(N_f\) \D5};
          \draw[fill=white] (2.5,5) circle (5pt);
          \draw (2.5,5) node[cross=4pt]{};
          \draw (0,5.5) node[]{\NS5};

          \draw (0,3) -- (2.5,3);
          \draw (1.2,2.5) node[]{\dots};
          \draw[fill=black] (0,3) node{} circle (5pt);
          \draw (1.9,3.5) node[]{\(N_f\) \D5};
          \draw[fill=white] (2.5,3) circle (5pt);
          \draw (2.5,3) node[cross=4pt]{};
          \draw (0,3.5) node[]{\NS5};

          \draw (0,1) -- (2.5,1);
          \draw (1.2,0.5) node[]{\((N_f - r_n)\) \D3};
          \draw[fill=black] (0,1) node{} circle (5pt);
          \draw (1.9,1.5) node[]{\(N_f\) \D5};
          \draw[fill=white] (2.5,1) circle (5pt);
          \draw (2.5,1) node[cross=4pt]{};
          \draw (0,1.5) node[]{\NS5};

          \draw (1,-.5) node[]{(c)};
        \end{footnotesize}
      \end{scope}

    \end{tikzpicture}
\end{center}
\caption{Duality in the \ac{kss} model
}
\label{figKSSduality}
\end{figure}
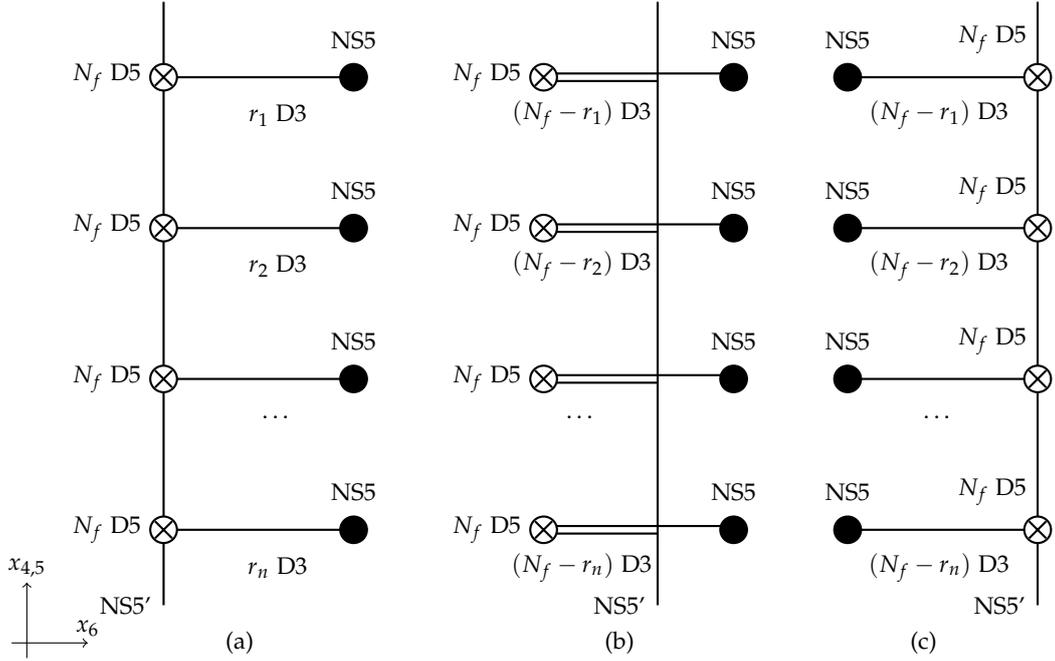

In the brane description we modify the \textsc{sqcd} analysis by introducing a 
set of $n$ \NS5--branes instead of one.
This induces the superpotential (\ref{WXk}).
The electric theory is represented on the left-hand side of Figure~\ref{figKSS}.
The different steps of the duality are shown in~Figure~\ref{figKSSduality}.
First we separate the  $n$ \NS{}--branes, and this procedure generates the 
superpotential (\ref{polyX}).
This separation breaks the gauge symmetry to $\prod_{i=1}^{n} U(r_i)$.
In the brane picture this is reflected by the separation of the $N_c$ \D3--branes along $(8,9)$.
Finally we move the \NS{}s and the \D3--branes in $(8,9)$ and come back to 
the original stack. The final configuration is depicted on the right-hand side of Figure~\ref{figKSS}.

\begin{figure}
  \begin{center}
    \begin{tikzpicture}
      \begin{scope}[shift={(-2,1)}]
        \begin{footnotesize}
          \draw[->] (0,3) -- (1,3);
          \draw[->] (.2,2.8) -- (.2,3.8);
          \draw (1,3.2) node[]{\(x_6\)};
          \draw (.2,4) node[]{\(x_{4,5}\)};
        \end{footnotesize}
      \end{scope}

      \begin{scope}[thick]
        \begin{footnotesize}

          \draw (0,1) -- (0,5);
          \draw (-0.5,1) node[]{\NS5'};
          \draw (0,3) -- (4,3);
          \draw (1.5,2.5) node[] {\(N_c\) \D3};
          \draw[fill=black] (4,3) node{} circle (5pt);
          \draw (4,3.5) node{\(n\) \NS5};
          \draw[fill=white] (0,3) circle (5pt);
          \draw (0,3) node[cross=4pt]{};
          \draw (-0.75,3) node[]{\(N_f\) \D5};
        \end{footnotesize}
      \end{scope}
      \begin{scope}[shift={(6,0)},thick]
        \begin{footnotesize}
          \draw (4,1) -- (4,5);
          \draw (3.5,1) node[]{\NS5'};
          \draw (0,3) -- (4,3);
          \draw (1.5,2.5) node[]{\((n N_f - N_c)\) \D3};
          \draw[fill=black] (0,3) node{} circle (5pt);
          \draw (0,3.5) node[]{\(n\) \NS5};
          \draw[fill=white] (4,3) circle (5pt);
          \draw (4,3) node[cross=4pt]{};
          \draw (4.75,3) node[]{\(N_f\) \D5};
        \end{footnotesize}
      \end{scope}
    \end{tikzpicture}
  \end{center}
  \caption{Electric and magnetic version of \ac{kss} from the brane picture.
}
  \label{figKSS}
\end{figure}
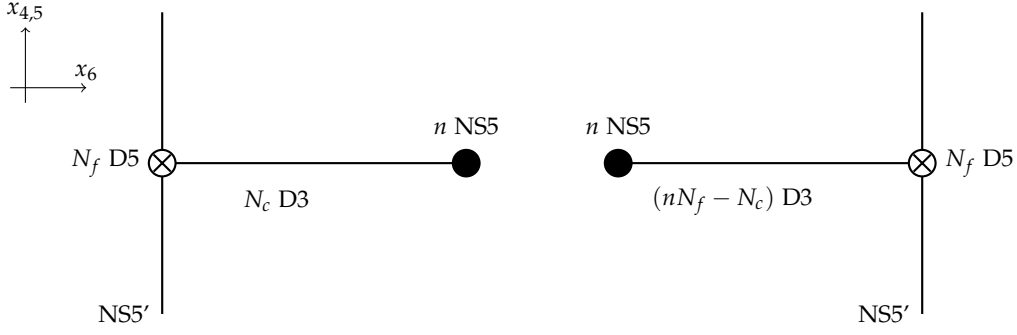

After T--duality along $x_3$ the \D4 and the \D6--branes become \D3s and \D5s respectively, while 
the \NS{}--branes are unchanged.
The theory at finite radius is represented in Figure~\ref{figKSSduality}.
Let us consider first the electric configuration (Figure~\ref{figKSSduality} (a)).
It consists of a set of decoupled \textsc{sqcd} models, each with $N_f$ flavors. There are $n$ of
those sectors, each with gauge group $U(r_i)$, with $\sum_{i=1}^{n} r_i=N_c$.
There is no adjoint matter anymore.

The superpotential (\ref{brokenetaadj}) is generated.
The same procedure can be implemented in the magnetic case.
In this way we recover the duality with the $\eta$--superpotential discussed above.
At this point we can flow to the \ac{kp} duality by separating the \D5--branes along $x_3$ as done
in the case of \textsc{sqcd}.
The final duality is obtained by reconstructing the stack of $n$ \NS{}--branes, \emph{i.e.} by
sending the superpotential (\ref{polyX}) to zero.

\section{Outlook}

In this section we outline further applications.
A straightforward generalization involves real gauge groups or matter in more complicated gauge representations. 
It would also be interesting to extend the analysis to cases with higher supersymmetry or to other dimensions.

\subsection{Orientifold planes}

Real gauge groups (symplectic or orthogonal)  and tensor matter (for example symmetric or antisymmetric representations) are obtained by adding 
orientifold planes.
Here we discuss the strategy for the brane reduction for both cases.

\paragraph{Real gauge groups.}
The reduction of Seiberg duality for $Sp(2 N_c)$ \textsc{sqcd} with $2N_f$ flavors has been discussed from
the field-theoretical point of view in~\cite{Aharony:2013dha}. The $SO(N_c)/O(N_c)$ cases have been studied in~\cite{Aharony:2013kma}.  
One can describe these theories in terms of the brane systems by adding \O6--planes or \O4--planes.
Different constructions are possible, we refer to~\cite{Giveon:1998sr} for reference.
The main idea is to study the reduction and the generation of the $\eta$--superpotential in the same 
way as above: one first compactifies $x_3$ and then performs a T--duality.
An $\eta$--superpotential is generated by the compactness of $x_3$
because of the \D1--strings stretched between the \D3--branes.
We leave the detailed analysis of this case and the matching with the results of~\cite{Aharony:2013dha,Aharony:2013kma} for future works.

\paragraph{Tensor matter.}

The orientifold projection can be applied to more complicated brane systems. One can for example
consider a stack of \D4--branes between two parallel \NS{}--branes. If there are an \NS'--brane and an orientifold plane
between the two \NS{}--branes we still have a unitary theory, but in
this case it includes tensor matter
(symmetric or antisymmetric with its own conjugate representation). One can also consider fundamental matter fields 
by including \D6--branes.
In this case there are Seiberg-dual phases, obtained by interchanging the \NS{} and \NS'{}--branes.
We refer to~\cite{Giveon:1998sr} for references concerning the brane realization of these dualities.
There are some cases in which the dual theory is s--confining, and the reduction with the \(\eta\)--superpotential
has been applied to these systems in~\cite{Csaki:2014cwa}. One can in principle describe the reduction of the
s--confining theories in terms of brane systems.

Note that the theory on the circle can have both a Coulomb branch and a Higgs branch. Extra massless fields may appear at the intersection point between these two branches.
In some regions of this moduli space an $\eta$--superpotential is generated, while in other 
regions it is not. It would be extremely interesting to reproduce this behavior at the level of the brane system.

\subsection{Higher supersymmetry}

Another natural generalization of the construction that we have presented consists in
cases with higher supersymmetry.
We can consider the system studied in Section~\ref{sec:Braneology}, but with the \NS'--brane rotated into an
\NS{}--brane resulting in a system with two parallel \NS{}--branes. This system has $\mathcal{N}=2$ supersymmetry in four dimensions.
By applying the same reduction discussed above one ends up with a three-dimensional system
with $\mathcal{N}=4$ supersymmetry.
This system has an $SO(4)$ global  R--symmetry corresponding to the rotations in \(4589\) \emph{i.e.} the $SU(2)_C \times SU(2)_H$ symmetry underlining three-dimensional mirror symmetry.
As discussed in~\cite{Razamat:2014pta} one can in general reduce four-dimensional dualities for class-S theories~\cite{Gaiotto:2009we}
to three dimensions.
Another possibility consists in studying the $\mathcal{N}=1$ dualities discussed in~\cite{Gadde:2013fma,Agarwal:2013uga}, 
obtained by coupling two copies of the theories of~\cite{Gaiotto:2009we}.

\subsection{Reduction to two dimensions}

Finally we can consider the reduction of four-dimensional Seiberg dualities to two dimensions.
In this case it is possible to switch on new parameters related to the compactification torus, \emph{i.e.} by introducing twistings~\cite{Hellerman:2011mv,Orlando:2011nc,Hellerman:2012zf,Orlando:2013yea} or magnetic fields~\cite{Kutasov:2013ffl}. The resulting gauge theories inherit these parameters as couplings for the fields, \emph{e.g.} twisted masses for the adjoints and fundamentals~\cite{Hellerman:2011mv}. We are thus lead to Seiberg-like dualities for families of \(\mathcal{N}=(1,1)\) or \(\mathcal{N}=(0,2)\) theories in two dimensions with matching parameters that can be read off directly from brane constructions analogous to the one that we have introduced in this paper. 
Such dualities have been observed directly from a two-dimensional perspective in the literature~\cite{Hanany:1997vm,Orlando:2010uu,Orlando:2010aj,Benini:2012ui}.

\section*{Acknowledgments}
It is a pleasure to thank Alberto Zaffaroni, Costas Bachas, Diego Redigolo, Mario Martone and Prarit Agarwal
for useful discussions and comments and Alberto Zaffaroni for insightful comments on the draft.
A.A. is funded by the European Research Council
(\textsc{erc}-2012-\textsc{adg}\_20120216) and  acknowledges support by \textsc{anr} grant
    13-\textsc{bs}05-0001. 
D.F. is \textsc{frs-fnrs} Charg\'e de Recherches. He acknowledges support by the \textsc{frs-fnrs}, 
by \textsc{iisn} - Belgium through conventions 4.4511.06 and 4.4514.08, by the Communaut\'e Francaise de Belgique 
through the \textsc{arc} program and by the \textsc{erc} through the SyDuGraM Advanced Grant. C.K. acknowledges support by \textsc{anr} grant 12-\textsc{bs}05-003-01 and by Enhanced Eurotalents, which is co-funded by \textsc{cea} and the European Commission.
D.O. would like to thank the theory group at \textsc{cern} for hospitality.

\setstretch{.95}

\printbibliography

\end{document}